\documentclass{article}
\usepackage{arxiv}
\usepackage[utf8]{inputenc} 
\usepackage[T1]{fontenc}    
\usepackage{hyperref}       
\usepackage{url}            
\usepackage{booktabs}       
\usepackage{amsfonts}       
\usepackage{nicefrac}       
\usepackage{microtype}      
\usepackage{lipsum}		
\usepackage{graphicx}
\usepackage{doi}

\usepackage{color}
\usepackage{tikz}
\usetikzlibrary{quantikz}
\usepackage{amssymb} 
\usepackage{caption}
\usepackage{subcaption}
\usepackage{tabularx}
\usepackage{ragged2e}

\newcommand*\bigstrut{%
  \vrule height.5\baselineskip depth.35\baselineskip width 0pt\relax}
  
\newcommand{\Hadamard}{\text{H}}
\newcommand{\Identity}{\text{I}}

\newcommand{\Ry}{\text{R}$_\text{y}$}
\newcommand{\Rz}{\text{R}$_\text{z}$}
\newcommand{\CX}{\text{CX}}
\newcommand{\CU}{\text{CU}}
\newcommand{\SX}{\text{SX}}

\title{Improved FRQI on superconducting processors and its restrictions in the NISQ era}

\author{ \href{https://orcid.org/0000-0002-9955-4809}{\includegraphics[scale=0.06]{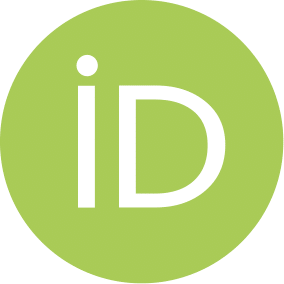}\hspace{1mm}Alexander Geng}\thanks{Corresponding author} \\
	Fraunhofer Institute for Industrial Mathematics ITWM\\ 
	Fraunhofer-Platz 1, 67663 Kaiserslautern\\            
	\texttt{alexander.geng@itwm.fraunhofer.de}\\
	\And
	\href{https://orcid.org/0000-0001-9126-3495}{\includegraphics[scale=0.06]{figs/orcid.png}\hspace{1mm}Ali Moghiseh}\\
	Fraunhofer Institute for Industrial Mathematics ITWM\\ 
	Fraunhofer-Platz 1, 67663 Kaiserslautern\\            
	\texttt{ali.moghiseh@itwm.fraunhofer.de}\\
	\And
	\href{https://orcid.org/0000-0002-8030-069X}{\includegraphics[scale=0.06]{figs/orcid.png}\hspace{1mm}Claudia Redenbach}\\
	University of Kaiserslautern\\ 
	Gottlieb-Daimler-Straße 47, 67663 Kaiserslautern\\            
	\texttt{redenbach@mathematik.uni-kl.de}\\
	\And
	\href{https://orcid.org/0000-0003-4903-3180}{\includegraphics[scale=0.06]{figs/orcid.png}\hspace{1mm}Katja Schladitz} \\
	Fraunhofer Institute for Industrial Mathematics ITWM\\ 
	Fraunhofer-Platz 1, 67663 Kaiserslautern\\            
	\texttt{katja.schladitz@itwm.fraunhofer.de}\\
}

\renewcommand{\shorttitle}{Improved FRQI}

\hypersetup{
pdftitle={Improved FRQI on superconducting processors and its restrictions in the NISQ era},
pdfsubject={q-bio.NC, q-bio.QM},
pdfauthor={Alexander Geng, Ali Moghiseh, Claudia Redenbach, Katja Schladitz},
pdfkeywords={Quantum image processing, Flexible Representation of Quantum Images, simulation, feasibility study, IBM Quantum Experience, real backend},
}

\begin{document}
\maketitle

\begin{abstract}
In image processing, the amount of data to be processed grows rapidly, in particular when imaging methods yield images of more than two dimensions or time series of images. Thus, efficient processing is a challenge, as data sizes may push even supercomputers to their limits. 
Quantum image processing promises to encode images with logarithmically less qubits than classical pixels in the image. In theory, this is a huge progress, but so far not many experiments have been conducted in practice, in particular on real backends. Often, the precise conversion of classical data to quantum states, the exact implementation, and the interpretation of the measurements in the classical context are challenging. We investigate these practical questions in this paper. In particular, we study the feasibility of the Flexible Representation of Quantum Images (FRQI). Furthermore, we check experimentally what is the limit in the current noisy intermediate-scale quantum era, i.e. up to which image size an image can be encoded, both on simulators and on real backends. Finally, we propose a method for simplifying the circuits needed for the FRQI. With our alteration, the number of gates needed, especially of the error-prone controlled-NOT gates, can be reduced. As a consequence, the size of manageable images increases.
\end{abstract}

\keywords{Quantum image processing \and Flexible Representation of Quantum Images \and simulation \and feasibility study \and IBM Quantum Experience \and real backend}

\section{Introduction}
With the help of increasingly powerful supercomputers, we are nowadays able to solve image processing problems that have been unsolvable just a few years ago. Especially in artificial intelligence and machine learning, higher computing power and the rise of deep convolutional neural networks recently enabled a leap forward. For some problems, often referred to as NP-hard problems, an exact solution using classical computers is not possible. Some of these NP-hard problems, however, can be solved efficiently on quantum computers, such as the integer factorization or discrete logarithm problem \cite{shor1999polynomial}. One goal of research in the area of quantum computing is to demonstrate quantum supremacy, that is, to show the superiority of a quantum computer over classical supercomputers for solving a complex problem \cite{preskill2012quantum,preskill2018quantum}.

In image processing, it has become increasingly important to process very large images, e.g. images bigger than one gigabyte, efficiently. A classical solution for this is to apply algorithms to a sub-division of the image and to subsequently merge the results. To avoid such splitting of the data and the associated edge effects, the transfer of the problem to the quantum world is promising.

In general, quantum image processing consists of three parts: quantum image representation, quantum image processing algorithms, and quantum image measurement. In classical image processing, the focus is on the actual algorithm and its efficiency. In quantum image processing, it is of highest priority to convert classical images into quantum states. For this step, there is a variety of algorithms including the Qubit Lattice \cite{venegas2003storing}, the Real Ket quantum image representation \cite{latorre2005image}, and the Flexible Representation of Quantum Images (FRQI) \cite{le2011flexible}. These three methods are considered to be the fundamental image representation options and serve as basic building blocks of many algorithms and starting points for other quantum image representations, see e.g. \cite{sun2011multi,sun2013rgb,zhang2013neqr,zhang2015qsobel}.

In this paper, we take a closer look at FRQI. Often this method is only described in theory and not explicitly applied on real backends. Mostly, only the minimal image size of $2\times 2$ pixels is considered. Here, we show how larger images can be implemented. Furthermore, we explore the current limitations in the noisy intermediate-scale quantum (NISQ) era, in which the results of the quantum computer are affected by noise. On a circuit-based superconducting quantum computer of IBM \cite{ibm}, we investigate up to which size an image can currently (August 2021) be encoded and measured. We increase the possible image size by improving the implementation of Qiskit \cite{qiskit_short} based on an idea that is also beneficial for other image representations and algorithms. 

This paper is organized as follows. In Section~\ref{sec:implementation}, we first explain the encoding of classical data into quantum states using the FRQI implementation. A naive implementation together with two pre-processing approaches is presented in Section~\ref{sec:practical_implementation}. Additionally, we explain two approaches for recovering the classical data from the empirical probabilities of the quantum computing results. In Section~\ref{sec:modification_mary}, we introduce an alternative to the naive implementation which has advantages especially for larger images. In Section~\ref{sec:extension}, we propose a way to extend both implementations to larger images.
In Section~\ref{section:QC_environment}, we present the experimental setup for evaluating the accuracy of the implementation on actual quantum computers. The results and the current maximum possible image size are discussed in Section~\ref{sect:results}. Section~\ref{sect:conclusion} concludes the paper.

\section{FRQI image representation}
\label{sec:implementation}
For representing a $2^n\times 2^n$ pixel gray value image, just $2n+1$ qubits are needed - $2n$ qubits for the position and one qubit for the gray value information of the pixels. In the basis states, each qubit is in a two-dimensional state, either $\ket{0}$ or $\ket{1}$. A combination of $2n$ qubits via the tensor product $\otimes$ yields $2^{2n}$-dimensional computational basis states $\ket{i}$ for the position information. For simplicity, the basis states $\ket{i}$ are named by the decimal number $i$ corresponding to the vector of zeros and ones read as binary number. 

\begin{figure}[tb]
    \begin{minipage}{0.2\textwidth}
        \centering
        \includegraphics[width=0.9\linewidth,keepaspectratio=true]{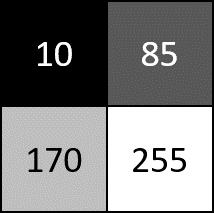}
    \end{minipage}
    \begin{minipage}{0.2\textwidth}
        \centering
        \includegraphics[width=0.9\linewidth,keepaspectratio=true]{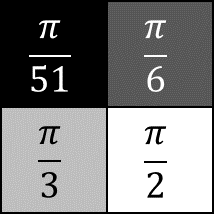}
    \end{minipage}
    \hspace{0.2cm}
    \begin{minipage}{0.5\textwidth}
        \begin{equation*}
            \begin{split}
                \ket{Img(\theta)} = &\frac{1}{2}[(\cos(\pi/51)\ket{0}+\sin(\pi/51)\ket{1})\otimes \ket{00}\\
                &+(\cos(\pi/6)\ket{0}+\sin(\pi/6)\ket{1})\otimes \ket{01}\\
                &+(\cos(\pi/3)\ket{0}+\sin(\pi/3)\ket{1})\otimes \ket{10}\\
                &+(\cos(\pi/2)\ket{0}+\sin(\pi/2)\ket{1})\otimes \ket{11}]\\
            \end{split}
        \end{equation*}
        \null
        \par\xdef\tpd{\the\prevdepth}
    \end{minipage}
\caption{$2\times 2$ pixel gray value image, transformation to angles in the range $[0,\pi/2]$, and the resulting quantum state representation.}
\label{fig:geng_alexander:image_in}
\end{figure}

The gray value information of the pixels is encoded by
\begin{equation}\label{equ:frqi_color_state}
    \ket{c_i}=\cos(\theta_i)\ket{0}+\sin(\theta_i)\ket{1} \text{ and } \theta_i\in\left[0,\frac{\pi}{2}\right],
\end{equation}
where $\theta=\left(\theta_0, \theta_1, \dots, \theta_{2^{2n}-1}\right)$ is a vector of angles encoding the gray values of each pixel. Section~\ref{section:preprocessing} provides more details on how to convert gray values into angles. In total, the two states for position and gray value information of every pixel are converted into the FRQI state by the formula
\begin{equation}\label{equ:frqi_state}
    \ket{Img(\theta)}=\frac{1}{2^n}\sum_{i=0}^{2^{2n}-1}\ket{c_i}\otimes\ket{i}.
\end{equation}
Figure~\ref{fig:geng_alexander:image_in} shows a $2\times 2$ sample gray value image and its FRQI state.

In quantum computing, the qubits are usually initialized in well-prepared states. In Qiskit \cite{qiskit_short}, the $\ket{0}$ state is the initial one for all qubits. Thus, the initial state of a quantum circuit for FRQI is $\ket{0}^{\otimes 2n+1}$. To convert this initial state into the FRQI state from Equation~\eqref{equ:frqi_state}, we apply single- and multi-qubit gates. The polynomial preparation theorem \cite{le2011flexible} shows, that the desired FRQI state can be constructed efficiently in two steps. First, all $2n$ position qubits have to be superposed to cover all $2^{2n}$ pixels in the classical image. This is achieved by using $2n$ Hadamard (\Hadamard) gates in parallel for the position qubits. That is, we apply the tensor product $I \otimes H^{\otimes 2n}$ to the initial state
\begin{equation}
    \ket{H}=(I \otimes H^{\otimes 2n})\ket{0}^{\otimes2n+1}=\frac{1}{2^n}\ket{0}\otimes\sum_{i=0}^{2^{2n}-1}\ket{i}.
\end{equation}
In the second step, controlled rotation gates are applied
\begin{equation}
    \ket{Img(\theta)}=\left(\prod_{i=0}^{2^{2n}-1} R_i\right)\ket{H},
\end{equation}
where
\begin{equation}\label{equ:controlled_rotation}
    R_i=\left(I \otimes \sum_{j=0,j\neq i}^{2^{2n}-1}\ket{j}\bra{j}\right)+R_y(2\theta_i)\otimes \ket{i}\bra{i}.
\end{equation}
Here, $\bra{\cdot}$ represents the adjoint of $\ket{\cdot}$ and $\ket{\cdot}\bra{\cdot}$ is the outer product. The first part of Equation~\eqref{equ:controlled_rotation} describes the control qubits and the second part the rotation using the \Ry-gate \cite{nielsen2000quantum} defined as
\begin{equation}\label{equ:ry-gate}
    R_y(\theta_i)=\left(\begin{array}{ccr}
        \cos(\theta_i/2) && -\sin(\theta_i/2) \\
        \sin(\theta_i/2) && \cos(\theta_i/2)
    \end{array}\right).
\end{equation}
Each multi-controlled operation, like the controlled rotation above, can be decomposed into single-qubit and two-qubit operations \cite{nielsen2000quantum}. This allows us to implement such operations on quantum computers. On these, all operations have to be transformed into basis gates, which for IBM's backends are currently identity (\Identity), NOT (X), square-root NOT (\SX), rotation (\Rz), and controlled-NOT (\CX) gates \cite{ibm}. Typically, an operation can be decomposed in several ways. The decomposition chosen is decisive for the number of gates required and the resulting noise level. 

\section{Practical implementation of FRQI}\label{sec:practical_implementation}
\subsection{Pre-processing: Converting gray values into angles}\label{section:preprocessing}
For representing an image on the quantum computer with FRQI, the pixel values have to be converted into angles $\theta_i\in \left[0,\frac{\pi}{2}\right]$ (see Equation~\eqref{equ:frqi_color_state}). Here, we only consider 8-bit gray value images with input pixel values $v_{in, i}\in[0,255]$. The first and obvious approach for obtaining the angle representation is a linear transformation. The pixel values are converted to fall into the range $\left[0,\frac{\pi}{2}\right]$ via
\begin{equation}\label{equ:normalization}
    \theta_i= \frac{v_{in,i}}{255}\cdot\frac{\pi}{2}.
\end{equation}

Alternatively, we follow \cite{co2021quantum} and use the transformation
\begin{equation}\label{equ:theta_encoding}
    \theta_i=\arcsin\left(\frac{v_{in,i}}{255}\right),
\end{equation}
to get angles in the required interval $\left[0,\frac{\pi}{2}\right]$. Of course, the arcsin in Equation~\eqref{equ:theta_encoding} can also be replaced by an arccos to achieve angles in the required range. 

\subsection{MCRY-implementation}
Independent of the two transformation variants, the multi-controlled rotation gates from Equation~\eqref{equ:controlled_rotation} must be converted into gates, which are executable on a real quantum computer.
In Qiskit, the Multi-Controlled Y rotation gates (MCRY) are converted by some standard routines of Qiskit \cite{qiskit_short}. The MCRY-gates only apply a rotation to the gray value qubit if all position qubits are in state $\ket{1}$. Subsequently, the result of the rotation has to be transferred to the correct position qubit state by application of X-gates. This allows to encode the next pixel using state $\ket{1}$ of all position qubits. 

\begin{figure}[tb]
\begin{subfigure}[tb]{\linewidth}
  \begin{minipage}{.305\linewidth}
    \includegraphics[width=0.99\textwidth]{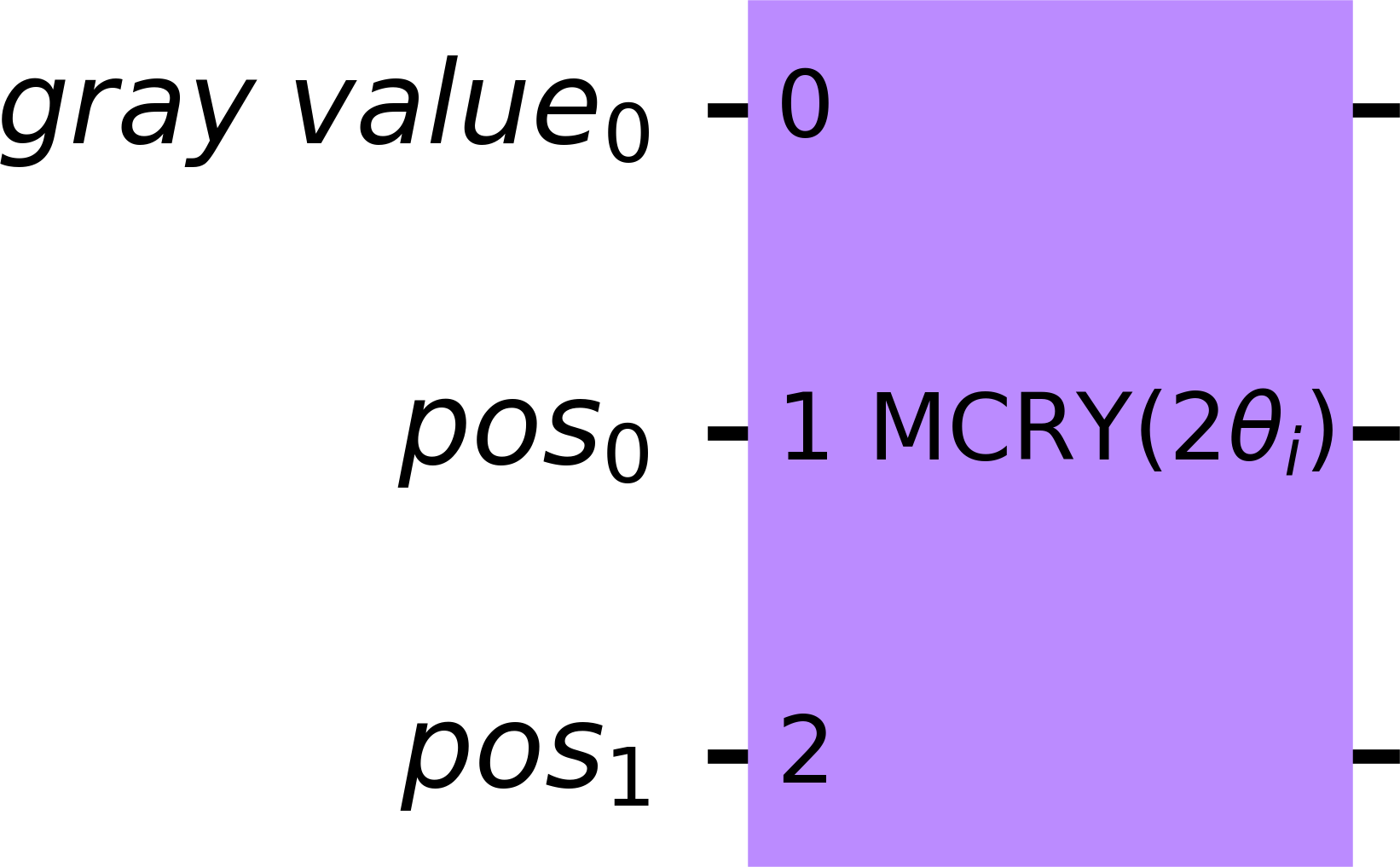}
  \end{minipage}%
  \begin{minipage}{.1\linewidth}
    \vspace{-0.5cm}
    \begin{eqnarray*}
       \hspace{0.5cm}\equiv
    \end{eqnarray*}
  \end{minipage}%
  \begin{minipage}{.595\linewidth}
    \includegraphics[width=0.99\textwidth]{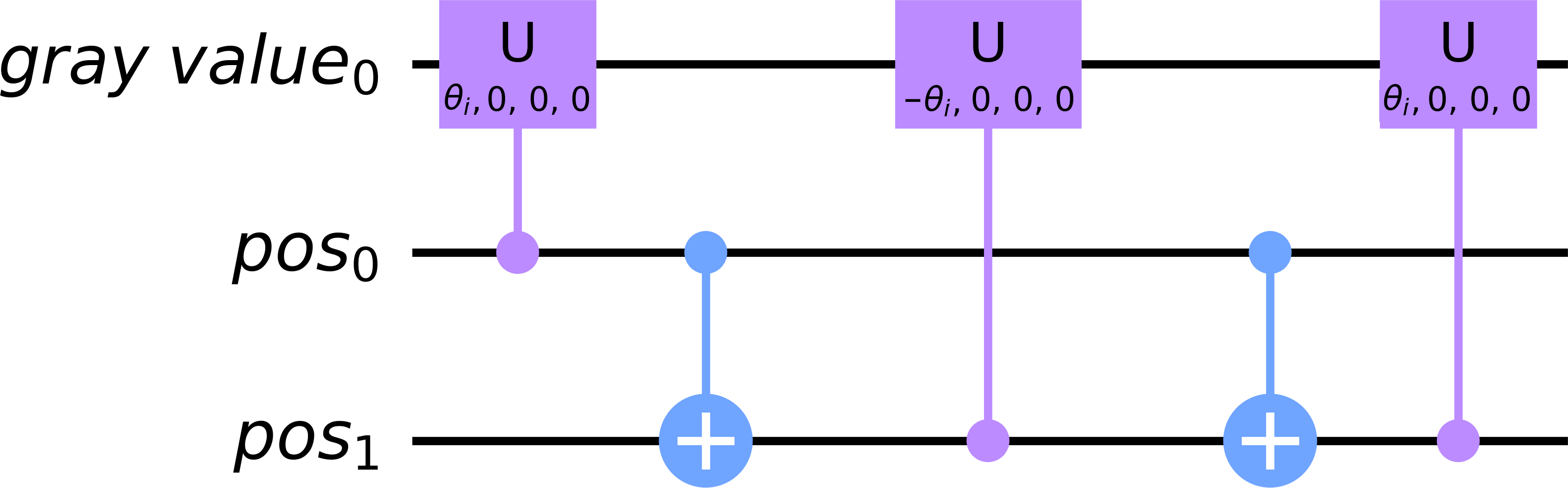}
  \end{minipage}
  \caption{MCRY-gate for an arbitrary angle $\gamma_i=2\theta_i$ and its decomposition into \CX- and \CU-gates. The \CU-gates must be converted further into the basis gates. In total, $8$ \CX-, $12$ \SX-, and $12$ \Rz-gates are needed for one MCRY-gate without considering the coupling map (the number of \CX-gates may increase due to additional SWAP-gates).\vspace{0.5cm}}
  \label{fig:geng_alexander:mcry_gate}
\end{subfigure}
\begin{subfigure}[tb]{\linewidth}
	\centering
	\includegraphics[width=0.99\textwidth]{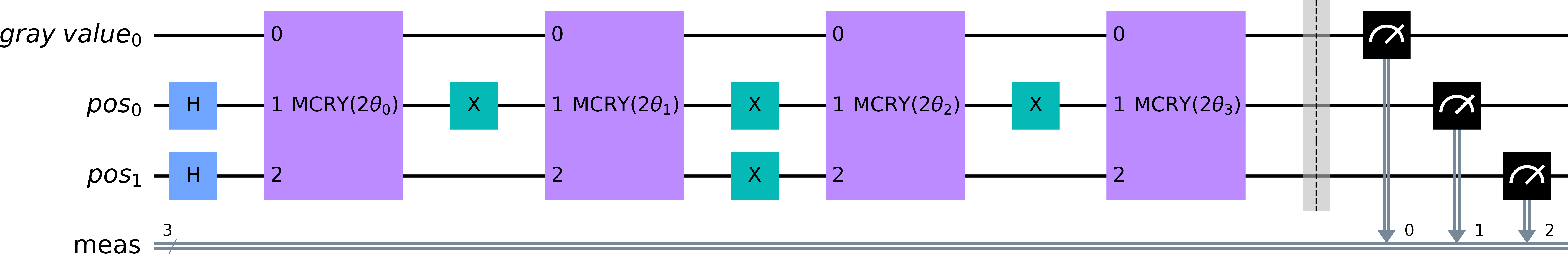}
	\caption{Circuit for an arbitrary $2\times 2$ gray value image using the MCRY-implementation. The pixel values are converted into angles $\theta_i$ for $i\in\{0,1,2,3\}$. Hadamard gates are used for superposition, X-gates for rotation to the desired state. There is one MCRY-gate for each pixel value.}
	\label{fig:geng_alexander:mcry_circuit}
\end{subfigure}
\caption{MCRY-gate and MCRY-implementation for a $2\times 2$ image.}
\end{figure}

One MCRY-gate is applied per input pixel. An example for a decomposition of one MCRY-gate with an arbitrary angle $\gamma_i=2\theta_i$ is shown in Figure~\ref{fig:geng_alexander:mcry_gate}. Here, the $2$ is needed as prefactor in Equation~\eqref{equ:ry-gate} to get the sine and cosine terms required in Equation~\eqref{equ:frqi_color_state}. 

A MCRY-gate is decomposed into two \CX-gates and three controlled-U-gates (\CU). The latter have to be converted into basis gates available on the real backends. That is, $2$ \CX-, $4$ \SX- and $4$ \Rz-gates per \CU-gate. Thus the number of gates, especially \CX-gates, increases further ($8$ \CX-, $12$ \SX-, and $12$ \Rz-gates for one MCRY-gate). The circuit for an arbitrary $2\times 2$ example using MCRY-gates is visualized in Figure~\ref{fig:geng_alexander:mcry_circuit}. Note that in the graphical representation of gates and circuits, the top or gray value qubit is always the target qubit. The remaining qubits are controls. In this implementation, the number of \CX-gates is high and requires the entanglement of all qubits. This means that we have to insert SWAP-gates for missing connections on the real backends.  

\subsection{Post-processing: Converting probabilities to gray values} \label{section:converting_pixel}
The state of the quantum computer after running the above circuit cannot be determined exactly. By a measurement, frequencies of the possible states are observed. 
The classical representation of the output image then has to be determined from the thus derived empirical probability distribution. 

Consider a state $i=c\otimes j$, where $c$ represents the state of the gray value qubit and $j$ the state of the position qubits.
Let $p_{j|c=\ket{0}}$ the conditional probability of observing state $j$ given that the gray value qubit is in state $\ket{0}$. Analogously, $p_{j|c=\ket{1}}$ is the conditional probability of observing $j$ with gray value qubit in $\ket{1}$. The output pixel value can be determined by
\begin{equation}\label{equ:decoding}
    v_{out,i}=\arccos\left(\sqrt{\frac{p_{j|c=\ket{0}}}{p_{j|c=\ket{0}}+p_{j|c=\ket{1}}}}\right)\cdot 255 \cdot \frac{2}{\pi}
\end{equation}
for the linear conversion as given in Equation~\eqref{equ:normalization}. The probabilities for the states with gray value qubit in state $\ket{0}$ are considered and normalized with the sum of the probabilities with gray value qubit in state $\ket{0}$ and $\ket{1}$ for each possible state. This ensures that the quotient in the first part of Equation~\eqref{equ:decoding} is in $[0,1]$ and the pixel values in the gray value range $[0,255]$. According to Equation~\eqref{equ:frqi_color_state}, arccos is applied and the pixel values are retrieved by inverting Equation~\eqref{equ:normalization}. 

In the second approach, by definition, we only need to consider the states where the gray value qubit is in state \ket{1} for retrieving the image. The output pixel values $v_{out,i}$ are given by
\begin{equation}\label{equ:v_out}
    v_{out,i}=2^{n}\cdot\sqrt{\bigstrut p_{j|c=\ket{1}}}\cdot 255.
\end{equation}
The factor $2^n$ cancels the equal state weighting in the FRQI definition \eqref{equ:frqi_state} and the factor $255$ converts the normalized angles into gray values. 

Obviously, states with gray value qubit in state \ket{0} are not considered in Equation~\eqref{equ:v_out}. Instead, the general weighting factor $2^n$ is used. When using a small numbers of measurements/shots and in the presence of noise, there is no way to ensure that $p_{j|c=\ket{0}}+p_{j|c=\ket{1}}=1/{2^{2n}}$ for $j\in\{0,2^{2n}-1\}$. This can lead to incorrectly weighted states and result in pixel values that are not in the gray value range $[0,255]$. Therefore, this approach is only useful if high number of measurements/shots can be performed.

Due to these problems, we choose the first approach and apply the linear transformation.

\section{Modification of the MCRY-implementation: MARY-implementation}\label{sec:modification_mary}
\begin{figure}[tb]
\begin{subfigure}[tb]{\linewidth}
  \begin{minipage}{.26\linewidth}
    \includegraphics[width=0.99\textwidth]{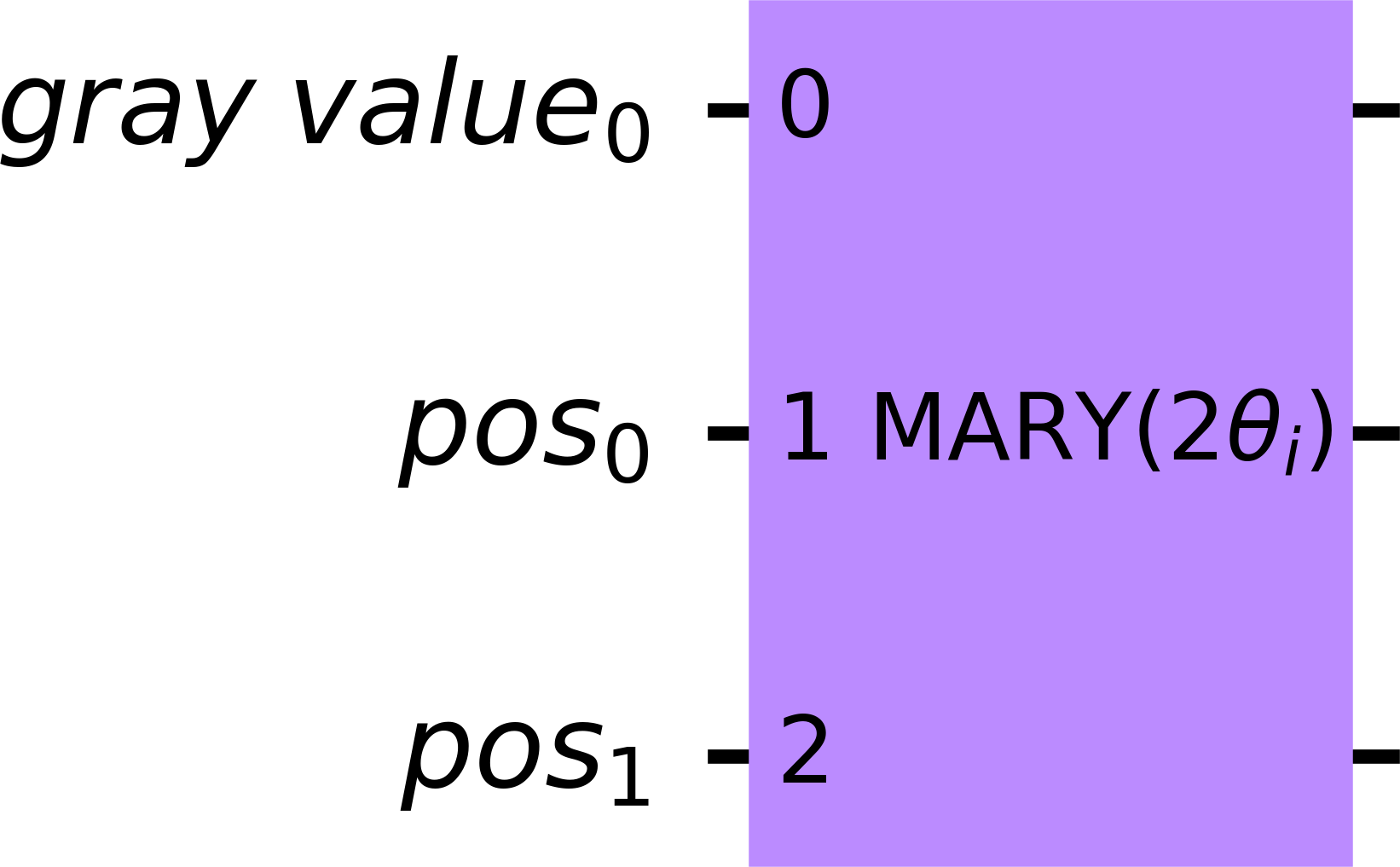}
  \end{minipage}%
  \begin{minipage}{.1\linewidth}
    \vspace{-0.5cm}
    \begin{eqnarray*}
       \hspace{0.5cm}\equiv
    \end{eqnarray*}
  \end{minipage}%
  \begin{minipage}{.64\linewidth}
    \includegraphics[width=0.99\textwidth]{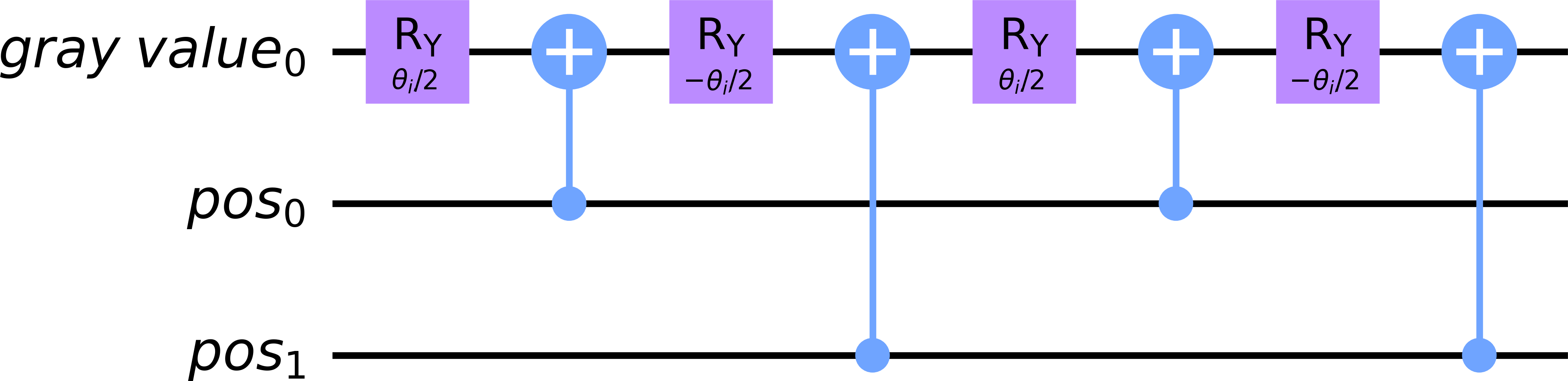}
  \end{minipage}
  \caption{MARY-gate for an arbitrary angle $\gamma_i=2\theta_i$ and its decomposition into \CX- and \Ry-gates. The \Ry-gates must be converted further into the basis gates. In total, $4$ \CX-, $8$ \SX-, and $8$ \Rz-gates are needed for one MARY-gate without considering the coupling map (the number of \CX-gates may increase due to additional SWAP-gates).\vspace{0.5cm}}
  \label{fig:geng_alexander:mary_gate}
\end{subfigure}
\begin{subfigure}[tb]{\linewidth}
	\centering
	\includegraphics[width=0.99\textwidth]{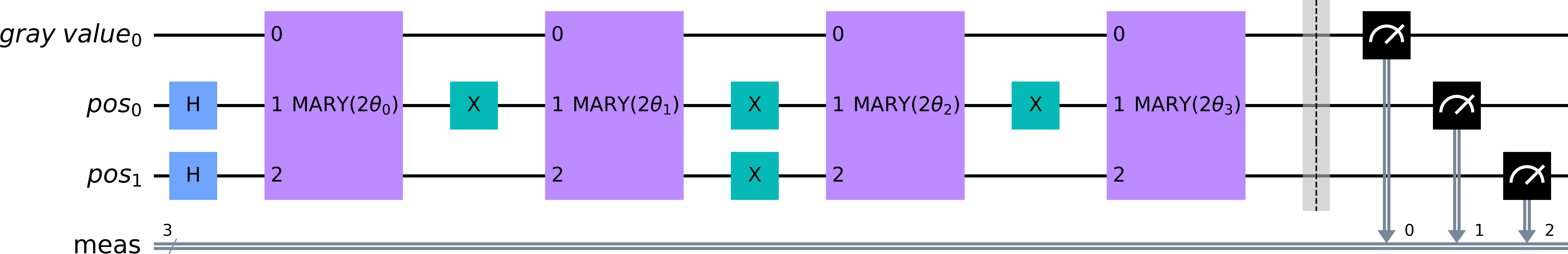}
	\caption{Circuit for an arbitrary $2\times 2$ gray value image using the MARY-implementation. Pixel values are converted into angles $\theta_i$, for $i\in\{0,1,2,3\}$. Hadamard gates are used for superposition, X-gates for rotation to the desired state. There is one MARY-gate for each pixel value.}
	\label{fig:geng_alexander:mary_circuit}
\end{subfigure}
\caption{MARY-gate and MARY-implementation for a $2\times 2$ image.}
\end{figure}

A more efficient method to implement the FRQI state for a $2\times 2$ image is inspired by \cite{co2021quantum}. The basic structure as well as the X- and Hadamard gates remain unchanged. However, instead of MCRY-gates we use MARY-gates for a part of the decomposition, leaving only \CX-gates or single-qubit Y rotation gates in the implementation. Figure~\ref{fig:geng_alexander:mary_gate} shows a MARY-gate with an arbitrary angle $\gamma_i=2\theta_i$, which is decomposed into four \Ry-gates and four \CX-gates. The \Ry-gates are further decomposed into \SX- and \Rz-gates, but the number of \CX-gates stays unchanged.

This way the number of error-prone \CX-gates is halved in comparison to the MCRY-implementation. Furthermore, the \CX-gates act only on mixed pairs of a gray value and a position qubit but not pairs of position qubits anymore. Consequently, not all three qubits have to be pairwise connected. Thus, the SWAP-gates required in the MCRY-implementation are avoided here. Figure~\ref{fig:geng_alexander:mary_circuit} shows the circuit for an arbitrary $2\times 2$ sample image. We only need to replace the MCRY- by MARY-gates.

\section{Extension to larger images}\label{sec:extension}
For larger images, the difference between the two implementations is more pronounced. In both cases, the position qubits must be entangled with the gray value qubit. In the MCRY-implementation, only the additional position qubits are added to the Qiskit MCRY-gate as control qubits. The decomposition of these multi-controlled rotation gates is left to the transpilation step and does not have to be adapted by the user. The effort to adapt the implementation to another image size is therefore very low. However, the required number of \CX-gates is very large, which increases the error rate, circuit depth, and execution time. 

\begin{figure}
  \begin{minipage}{.15\linewidth}
    \includegraphics[width=0.99\textwidth]{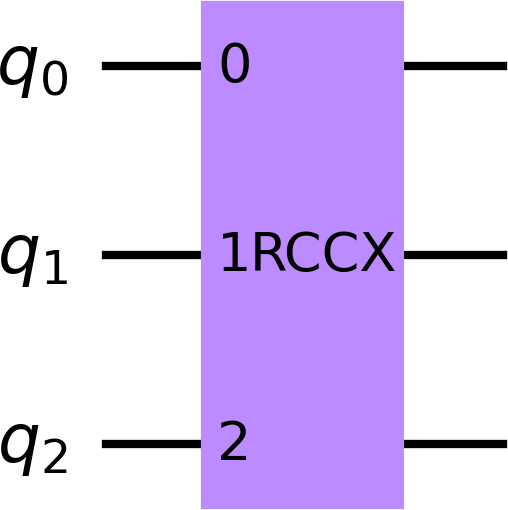}
  \end{minipage}%
  \begin{minipage}{.05\linewidth}
    \vspace{-0.5cm}
    \begin{eqnarray*}
       \equiv
    \end{eqnarray*}
  \end{minipage}%
  \begin{minipage}{.5\linewidth}
    \includegraphics[width=0.99\textwidth]{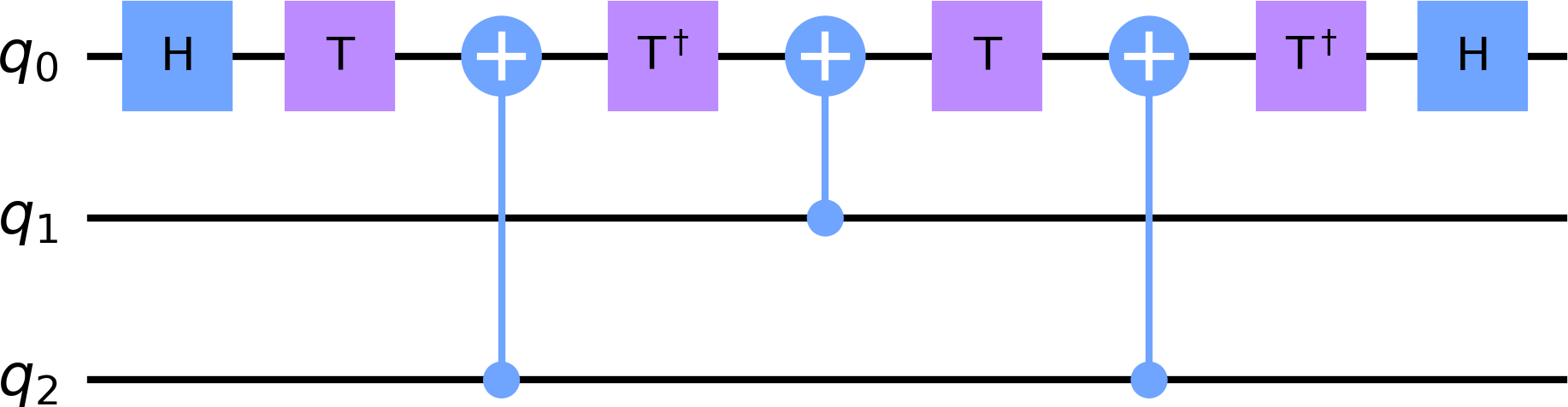}
  \end{minipage}
  \\\\\\\\
  \begin{minipage}{.15\linewidth}
    \includegraphics[width=0.99\textwidth]{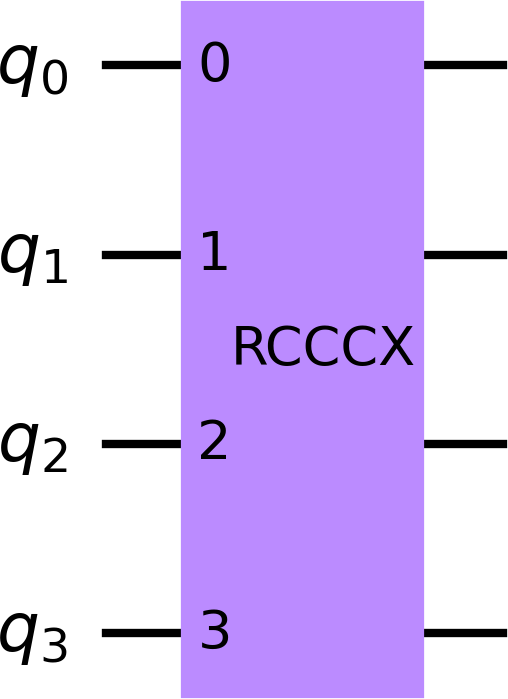}
  \end{minipage}%
  \begin{minipage}{.05\linewidth}
    \vspace{-0.5cm}
    \begin{eqnarray*}
       \hspace{0.05cm}\equiv
    \end{eqnarray*}
  \end{minipage}%
  \begin{minipage}{.8\linewidth}
    \includegraphics[width=0.99\textwidth]{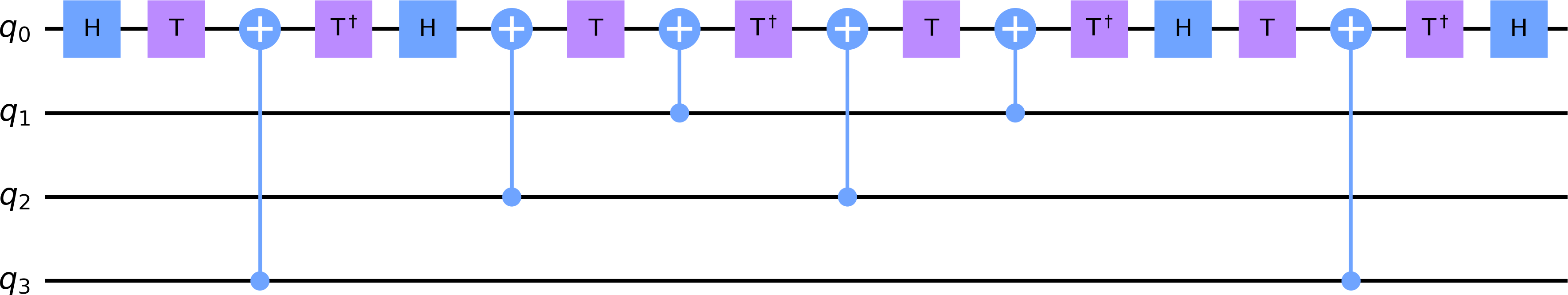}
  \end{minipage}
  \caption{Gates for entangling qubits instead of using multi-controlled \CX-gates. Their decompositions into single-qubit-gates and \CX-gates are shown on the right. Top: RCCX-gate; Bottom: RCCCX-gate.}
  \label{fig:geng_alexander:rccx_gate}
\end{figure}

To reduce the number of \CX-gates, we adapt the MARY-implementation for the larger images. We have to entangle the additionally required position qubits with the gray value qubit. Similar to \cite{co2021quantum}, we also use RCCX- or RCCCX-gates \cite{song2003simplified,maslov2016advantages} to entangle three or four qubits, respectively. These simplified versions of the naively implemented Toffoli gate or multi-controlled X-gate with three control qubits have the advantage of reducing the number of \CX-gates compared to the direct use of multi-controlled \CX-gates, see Figure~\ref{fig:geng_alexander:rccx_gate}. Hence, we only need $3$ or $6$ \CX-gates instead of $6$ or $14$, respectively. Note that the simplified versions of the gates have a different relative phase, so the elements in the matrix representation differ by a factor of $e^{i\pi \phi}, \phi\in \mathbb{R}$ \cite{maslov2016advantages}. Due to global phase invariance, this does not affect the results.

The main idea in constructing the circuits for larger images is to keep the operations on the gray value qubit the same as in \cite{co2021quantum} for all sizes, using X-gates to change the desired state and entangle the position qubits with the gray value qubit. We can entangle two qubits with a \CX-gate, three qubits with an RCCX-gate and four qubits with an RCCCX-gate. We have to combine these gates to entangle the $2n$ position qubits with the gray value qubit for a $2^n\times 2^n$ pixel gray value image. For that we have two ways: enlarging the MARY-gates by replacing \CX-gates with RCCX-, or RCCCX-gates or entangling the qubits before the MARY-gate and applying the corresponding gates symmetrically after the MARY-gate. In practice, one of the approaches or a combination of both are chosen depending on the image size.

\begin{figure}
    \includegraphics[width=0.99\textwidth]{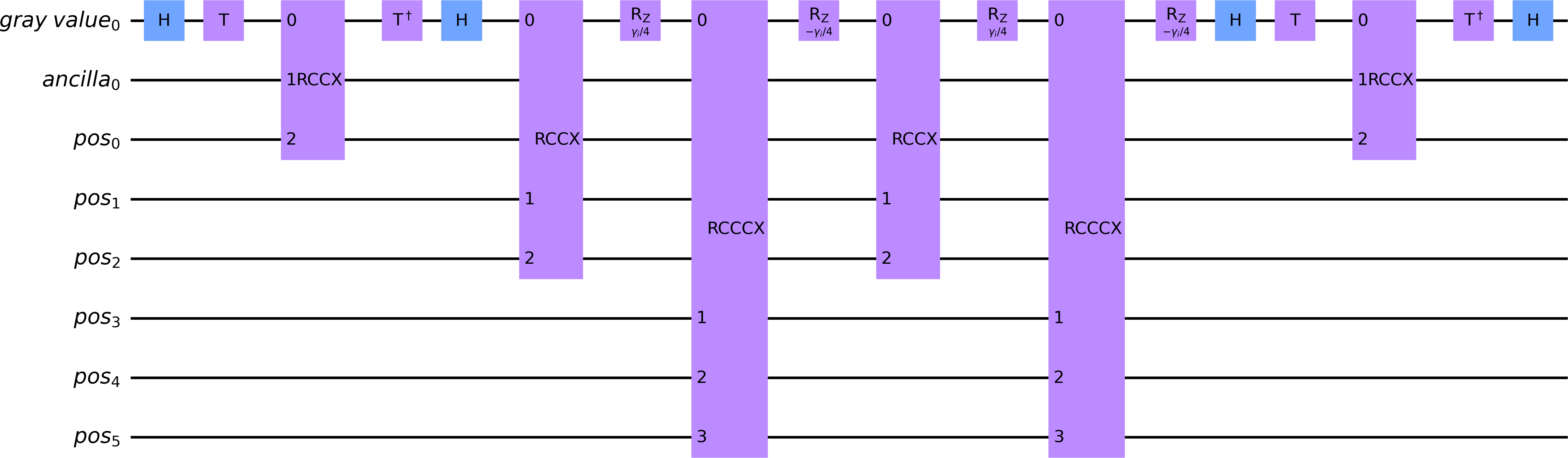}
    \caption{Decomposed MARY8-gate for an arbitrary angle $\gamma_i=2\theta_i$.}
    \label{fig:geng_alexander:mary8_gate}
\end{figure}

\begin{figure}[tb]
	\centering
	\includegraphics[width=0.8\textwidth]{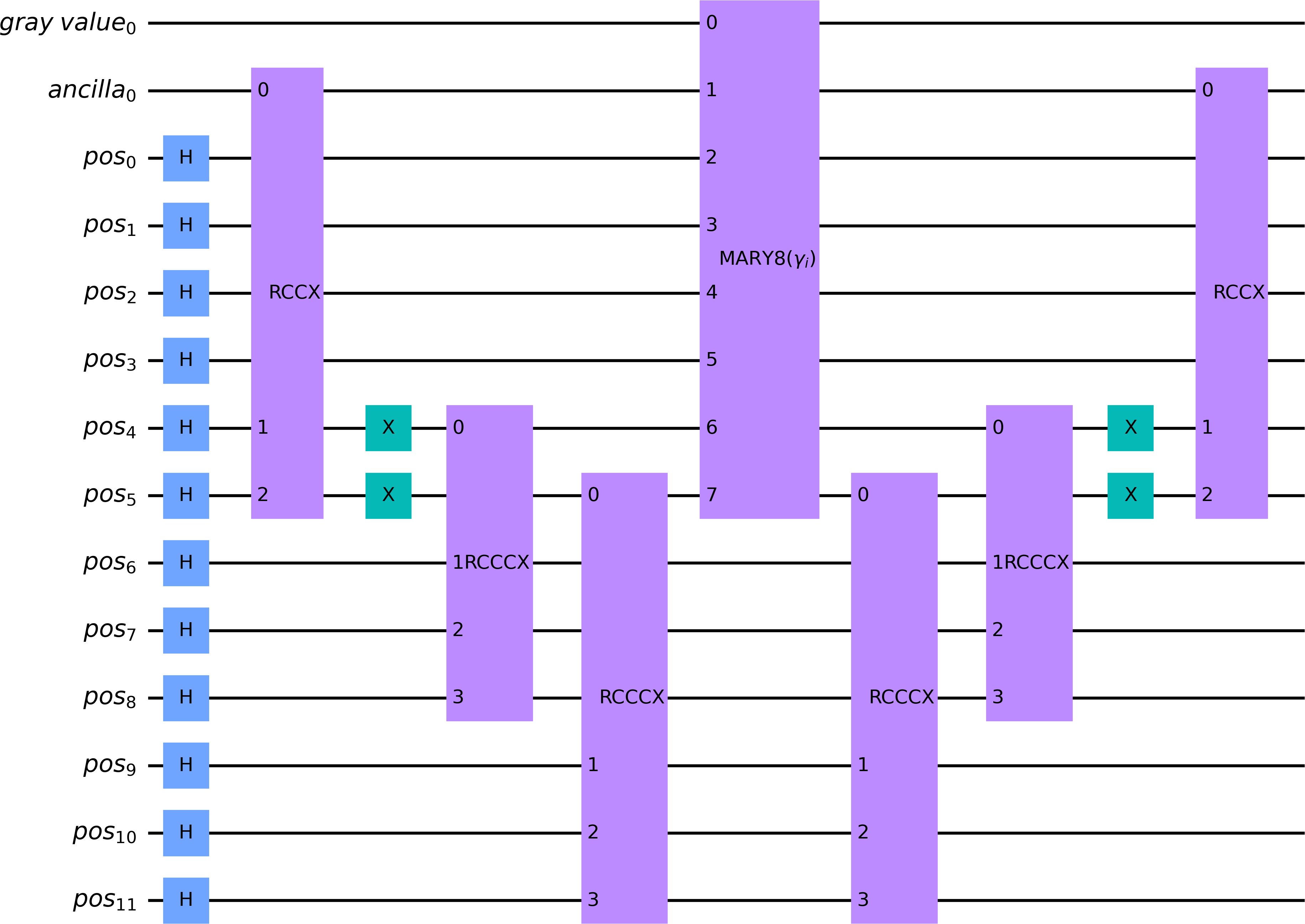}
	\caption{Circuit for one angle of a $64\times 64$ image. Hadamard gates are used for superposition, RCCX- and RCCCX-gates for entangling, and MARY8 for entangling and rotation. X-gates after the Hadamard gates 
	are not shown here.
	}
	\label{fig:geng_alexander:circuit64x64}
\end{figure}

Enlarging the MARY-gate allows to entangle a maximum of ten qubits using RCCCX-gates. An example of a MARY8-gate which  entangles eight qubits is shown in Figure~\ref{fig:geng_alexander:mary8_gate}. 
For images smaller than $32\times 32$, we can just increase the MARY-gate and keep the original workflow. However, for larger images, we have to both enlarge the MARY-gate and entangle outside the MARY-gate. Figure~\ref{fig:geng_alexander:circuit64x64} shows this combined way for a $64\times 64$ pixel gray value image. We entangle the position qubits before and symmetrically after the MARY8-gate. With the use of one ancilla qubit and one RCCX-gate, we can further combine the information from two position qubits. Using X-gates prepares the position qubits to receive information from other position qubits. This is achieved by two RCCCX-gates. Analogously, more RCCX- and RCCCX-gates can be used to encode images of other sizes.

Examples of possible circuits for image sizes $2^n\times2^n$, with $n=1,\dots, 9, 13$, are shown in the supplementary information in Section~\ref{section:supplementary}. This way, encoding of images of size $8.192 \times 8.192$ is conceivable by using one gray value qubit, three ancilla, and $26$ position qubits as visualized in the supplementary information in Figure~\ref{fig:geng_alexander:circuit8192}. The ancilla qubits serve only as storage qubits for entangled position qubits.

\section{Quantum computing environment used} \label{section:QC_environment}
Here, we describe our framework for realizing the methods from the previous section. It includes software, a classical computer, a quantum computer, and error models.

\begin{figure}[tb]
	\centering
	\begin{subfigure}[tb]{.2\linewidth}
        \includegraphics[width=0.99\textwidth]{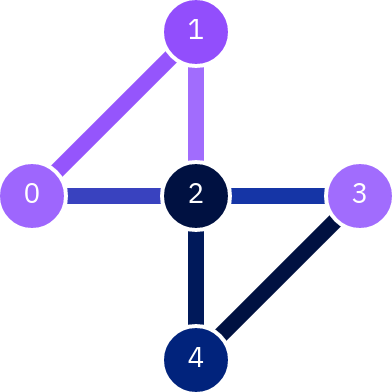}
        \caption{'ibmqx2'}
        \label{fig:geng_alexander:ibmqx2}
    \end{subfigure}
	\begin{subfigure}[tb]{.49\linewidth}
        \includegraphics[width=0.99\textwidth]{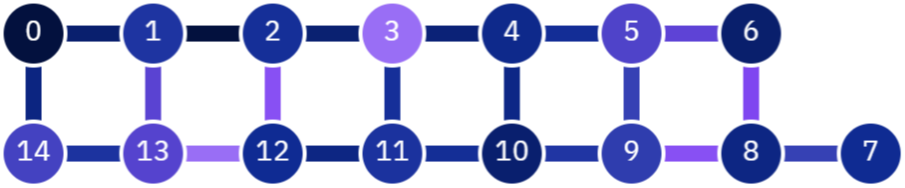}
        \caption{'ibmq\_16\_melbourne'}
        \label{fig:geng_alexander:melbourne}
    \end{subfigure}
    \begin{subfigure}[tb]{.29\linewidth}
        \includegraphics[width=0.99\textwidth]{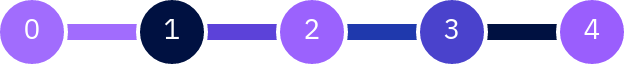}
        \caption{'ibmq\_santiago'}
        \label{fig:geng_alexander:santiago}
        \vspace{0.5cm}
        \includegraphics[width=0.99\textwidth]{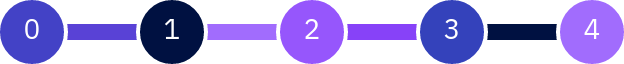}
        \caption{'ibmq\_manila'}
        \label{fig:geng_alexander:manila}
    \end{subfigure}
    \begin{subfigure}[tb]{.49\linewidth}
        \includegraphics[width=0.99\textwidth]{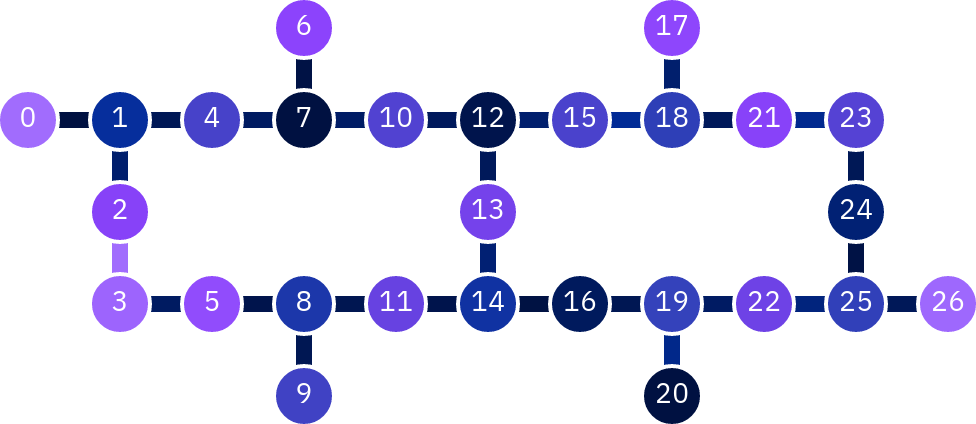}
        \caption{'ibmq\_toronto'}
        \label{fig:geng_alexander:toronto}
    \end{subfigure}
    \begin{subfigure}[tb]{.49\linewidth}
        \includegraphics[width=0.99\textwidth]{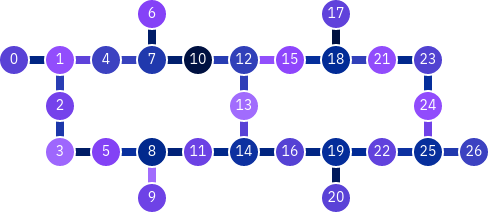}
        \caption{'ibmq\_ehningen'}
        \label{fig:geng_alexander:ehningen}
    \end{subfigure}
	\caption{Backends used in this paper. The qubit frequencies (points) and the \CX \,errors for the connections between the qubits (lines) are shown with various color values. Dark blue indicates small, purple high frequency/error.}
	\label{fig:geng_alexander:used_backends}
\end{figure}

We use the open-source software development kit Qiskit \cite{qiskit_short} for working with IBM's circuit-based superconducting quantum computers \cite{ibm}. Via cloud access, they provide a variety of systems, also known as backends, which differ in the number and performance of the qubits and their connectivity. In this paper, we use the backends 'ibmqx2', 'ibmq\_16\_melbourne', 'ibmq\_santiago', 'ibmq\_manila', 'ibmq\_toronto', and the German backend 'ibmq\_ehningen'. The corresponding coupling maps are shown in Figure~\ref{fig:geng_alexander:used_backends}.

The backends underlie external influences, so the values of the backends, like \CX \,error, readout error, or decoherence times, can change hourly. Calibration should diminish this effect, errors are however averaged over 24 hours. Typical average values for \CX \,error, readout error, decoherence times T1, T2, and frequency are shown in Table~\ref{tab:geng_alexander:error_rates}.

\begin{table}[tb]
    \caption{Typical average calibration data of the six chosen backends.}
    \label{tab:geng_alexander:error_rates}
    \centering
    \begin{tabular}{lccccc}
        \hline\noalign{\smallskip}
        Backend & \CX \,error & Readout error & T1 & T2 & Frequency \\
        &[\%] & [\%] & [$\mu \rm s$] & [$\mu \rm s$] & [GHz]\\
        \noalign{\smallskip}\hline\noalign{\smallskip}
        'ibmqx2'                & 1.90 & 5.66 & \phantom{1}50.03    & \phantom{1}34.74  & 5.187 \\
        'ibmq\_16\_melbourne'   & 3.40 & 5.64 & \phantom{1}56.44    & \phantom{1}55.53  & 4.980 \\
        'ibmq\_santiago'        & 0.60 & 1.59 & 148.27              & 106.61            & 4.767 \\
        'ibmq\_manila'          & 0.93 & 3.11 & 143.35              & \phantom{1}52.81  & 4.971 \\
        'ibmq\_toronto'         & 1.29 & 4.24 & \phantom{1}99.54    & 118.45            & 5.080\\
        'ibmq\_ehningen'         & 1.05 & 2.09 & \phantom{1}99.99    & 122.32            & 5.164\\
        \noalign{\smallskip}\hline
    \end{tabular}
\end{table}

Currently, all IBM quantum computers are affected by the errors just addressed. Measurement error mitigation is a way to reduce the errors, especially the readout error, and to improve the results \cite{Qiskit-Textbook_short}. The idea is to prepare all $2^n$ basis input states, execute them on the 'qasm\_simulator' with an error model, and to compute the probability of measuring basis states differing from the true input. Based on this, a backend specific calibration matrix is calculated which compares the desired and the measured states. Applying the inverse of this calibration matrix finally yields the improved results.

Two error models are used in the following. The Pauli error is applied to the measurement of the 'qasm\_simulator'. It consists in randomly flipping each bit in the output with a probability $p_{meas}$. In contrast, the depolarizing error model captures imperfections in operations. While processing, the state of any qubit is replaced by a completely random state with a probability of $p_{gate}$. For two qubit gates, like \CX-gates, the depolarizing error is applied independently to each qubit. For further details on these two error models and measurement error mitigation in general we refer to \cite{Qiskit-Textbook_short}.

In addition to quantum computers, a classical computer is needed for testing algorithms, reducing errors, and generating and storing the circuits before sending them to the quantum computer. The latter is the main limitation as circuits for encoding larger images need a lot of gates and have to be stored completely in the RAM. We use a computer with an Intel Xeon E5-2670 processor running at $2.60$ GHz, a total RAM of $64$ GB, and Red Hat Enterprise Linux 7.9. 

\section{Results and comparison of MCRY- and MARY-implementations}\label{sect:results}
\subsection{Sample image \texorpdfstring{$2\times 2$}{2 x 2}}

Our aim is to encode a classical image into a quantum state, measure the outcome, and compare the result with the classical input image. 

\begin{figure}[tb]
	\centering
	\includegraphics[width=0.99\textwidth]{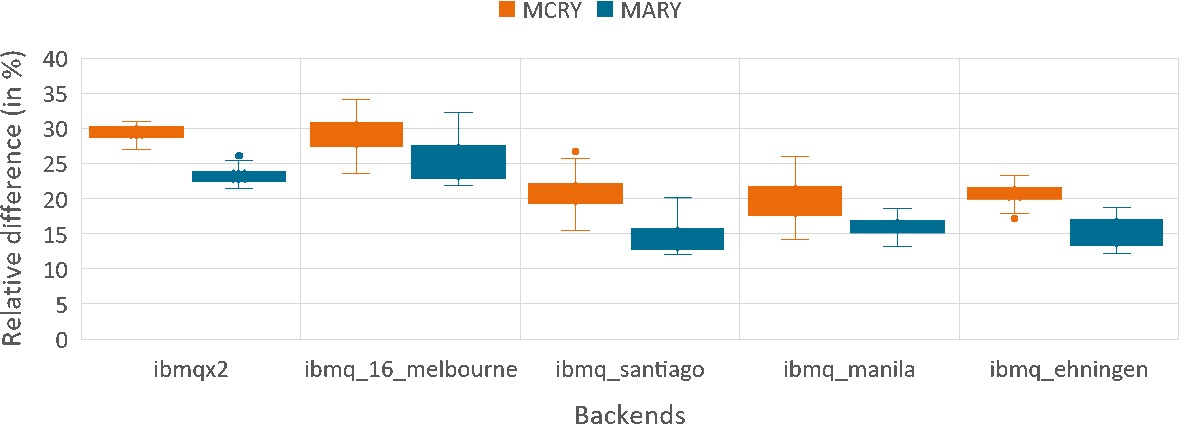}
	\caption{Box plots for relative difference between input image $[10,85,170,255]$ and output image from five different backends.}
	\label{fig:geng_alexander:boxplot_mcry_mary}
\end{figure}

We set the pixel values of the input image $image_{in,i}, i\in \{0,1,2,3\}$, to $[10,85,170,255]$ to cover the whole gray value range. We start at pixel value $10$, since $0$ is not generally interesting as only identity gates are needed for encoding it.
We measure deviation by the relative difference:
\begin{equation}
    diff_{rel}=\frac{1}{2^{2n}}\sum_{i=0}^{2^{2n}-1}|image_{out,i}-image_{in,i}|\cdot\frac{100}{255}
\end{equation}
for a general image size of $2^n\times2^n$. The term $image_{out}$ is the resulting image reconstructed from the measurements of the quantum computer, as described in Section~\ref{section:converting_pixel}. To compare the implementations, we ran the two circuits in one job on five different backends. This ensures that the same calibration is used for the two implementations and that differences in the outcome are not due to varying error rates or coherence times. Box plots for $42$ executions between June 7 and June 14, 2021 are shown in Figure~\ref{fig:geng_alexander:boxplot_mcry_mary}.

\begin{figure}[tb]
	\centering
	\includegraphics[width=0.99\textwidth]{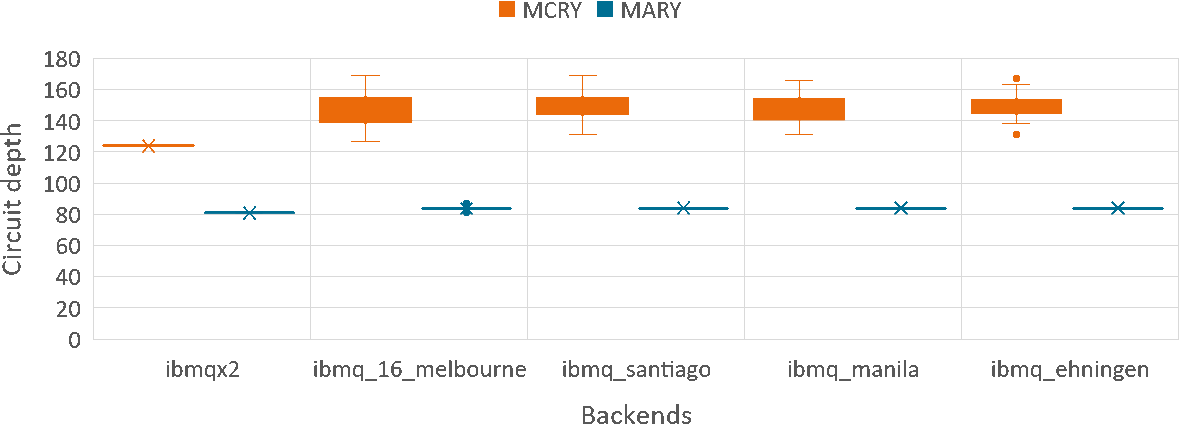}
	\caption{Box plots for circuit depth when using the sample image $[10,85,170,255]$.}
	\label{fig:geng_alexander:boxplot_circuit_depth_mcry_mary}
\end{figure}

\begin{figure}[tb]
	\centering
	\includegraphics[width=0.99\textwidth]{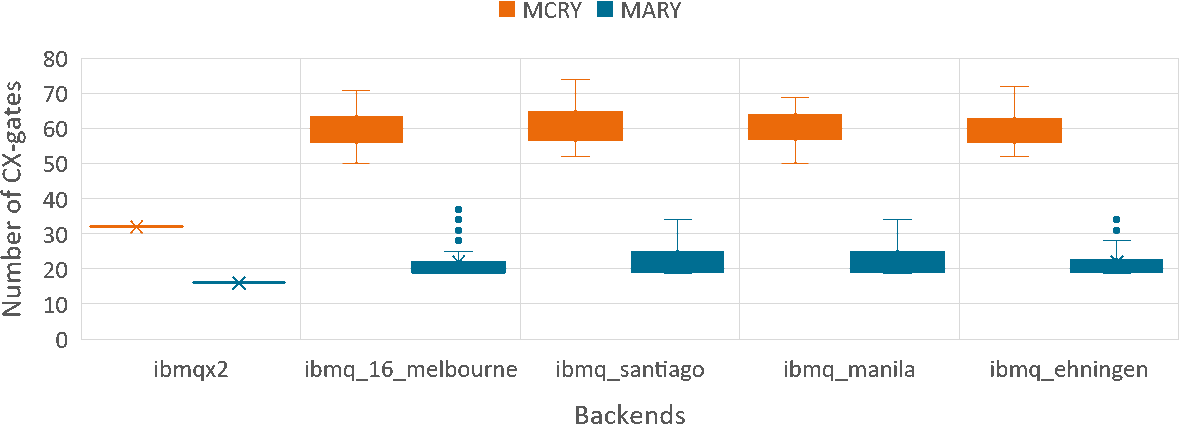}
	\caption{Box plots for number of \CX-gates when using the sample image $[10,85,170,255]$. }
	\label{fig:geng_alexander:boxplot_cx_gates_mcry_mary}
\end{figure}

Obviously, the MARY-implementations perform better than those with the MCRY-gates. This is due to the lower circuit depth (see Figure~\ref{fig:geng_alexander:boxplot_circuit_depth_mcry_mary}), especially the lower number of \CX-gates (see Figure~\ref{fig:geng_alexander:boxplot_cx_gates_mcry_mary}).

Differences between the five backends are clearly visible, too. Backend 'ibmqx2' has three fully-connected qubits and (like 'ibmq\_16\_melbourne') no heavy-hexagonal topology (see Figure~\ref{fig:geng_alexander:used_backends}). This yields the lower variance of the outcomes for this backend. This is even more apparent for the MCRY-implementation, where we can benefit from the high connectivity of the qubits in that no SWAP-gates have to be used. Newer backends use a purely heavy-hexagonal topology or segments thereof because it yields lower error rates than more connected topologies \cite{zhu2021hardware}. In Figure~\ref{fig:geng_alexander:boxplot_mcry_mary} this effect is visible in the lower relative difference of the younger backends 'ibmq\_santiago', 'ibmq\_manila', and 'ibmq\_ehningen' compared to the older 'ibmq\_16\_melbourne' and 'ibmqx2'.

As shown in Figure~\ref{fig:geng_alexander:boxplot_mcry_mary}, the MARY-implementation's results are more precise. Measurement error mitigation is a way to improve them further. 

Here, we pursue two ways to obtain the calibration matrix.
The first one, 'mitigation\_own', is executed on the 'qasm\_simulator' assuming a self-determined noise model. We follow the suggestions from \cite{Qiskit-Textbook_short} and assume Pauli-errors for the measurements and depolarizing errors for X- and \CX-gates. We set $p_{meas}=p_{gate}=10\%$,
which exceeds the actual error probabilities of the backends (see Table~\ref{tab:geng_alexander:error_rates}). 
This way, we can also account for other errors that cannot be incorporated directly, e.g., errors caused by the environment of the quantum computer. 

\begin{figure}[tb]
	\centering
	\includegraphics[width=0.99\textwidth]{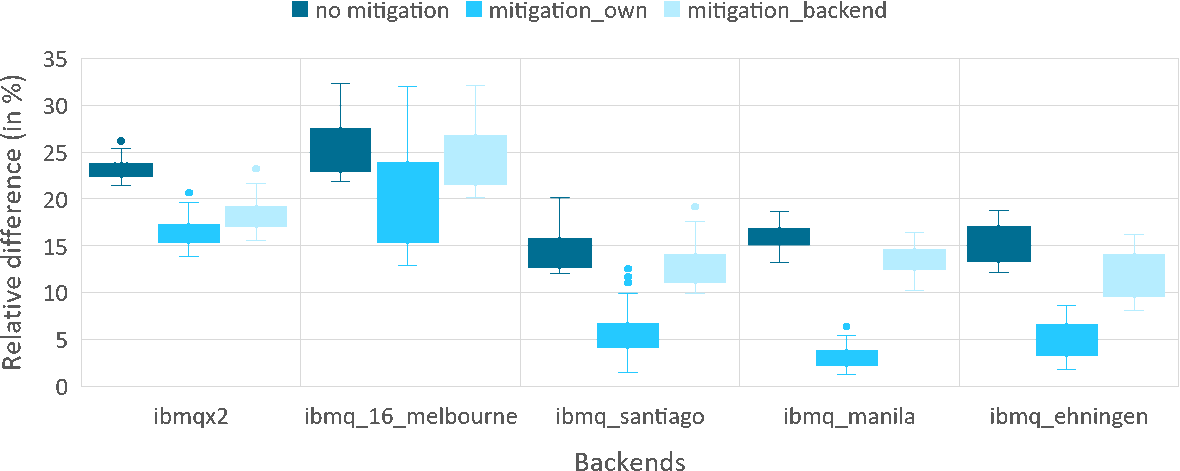}
	\caption{Box plots for relative difference between input image $[10,85,170,255]$ and output image from five different backends using the MARY-implementation. The backends 'ibmq\_santiago', 'ibmq\_manila', and 'ibmq\_ehningen' have heavy-hexagonal topology with lower error rates.}
	\label{fig:geng_alexander:boxplot_mcry_mary_mitigated}
\end{figure}

For the second approach, 'mitigation\_backend', we replace our noise model by the noise model from the Qiskit Aer Noise module \cite{qiskit_short} which is obtained via the command 'from\_backend(backend)'. It includes all error rates, coherence times, and the coupling map of the backend to that specific time. The outcome for the MARY-implementation is shown in Figure~\ref{fig:geng_alexander:boxplot_mcry_mary_mitigated}.

Both variants of measurement error mitigation reduce the relative difference for all tested backends. 
Figure~\ref{fig:geng_alexander:boxplot_mcry_mary_mitigated} shows, that 'mitigation\_own' works better. This is probably because 'mitigation\_backend' takes into account the actual daily qubit and gate errors of the backends but ignores other error sources such as crosstalk errors. In addition, other factors such as the environment of the quantum computer can also affect the result. With 'mitigation\_own', we circumvent this problem by fixing the error to $10\%$ which surely overestimates the true values. Changing or adjusting this value offers further potential for improving measurement error mitigation, but will not be considered further in this paper.

\begin{figure}[tb]
	\centering
	\begin{minipage}{0.16\linewidth}
	    \hspace{-0.4cm}
	    \centering
	    \resizebox{\linewidth}{!}{%
	    \begin{tabular}{c}
        \noalign{\smallskip}
        \includegraphics[height=1.5cm]{figs/image_in.png}\\
        Input image\\
        \end{tabular}}
    \end{minipage}
    \begin{minipage}{\linewidth}
            \centering
            \resizebox{\linewidth}{!}{%
            \begin{tabular}{c|ccc|ccc}
            \noalign{\bigskip}
            \multicolumn{7}{c}{\large MCRY-implementation}\\
            \noalign{\smallskip}
            \includegraphics[height=2.5cm]{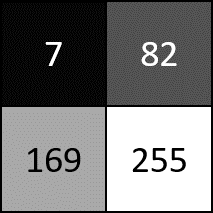} &  
            \includegraphics[height=2.5cm]{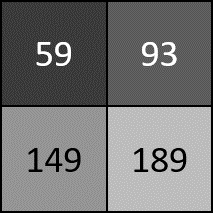}&
            \includegraphics[height=2.5cm]{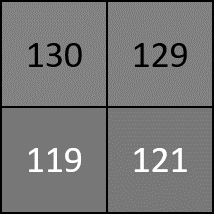} &
            \includegraphics[height=2.5cm]{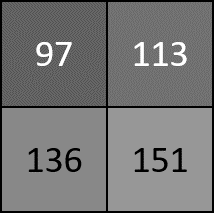} & \includegraphics[height=2.5cm]{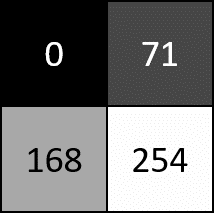}& 
            \includegraphics[height=2.5cm]{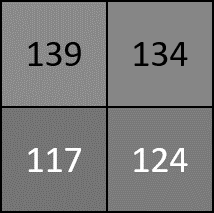}&
            \includegraphics[height=2.5cm]{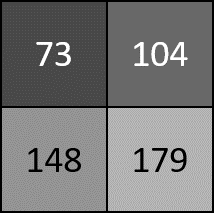}\\ \noalign{\bigskip}
            \multicolumn{7}{c}{\large MARY-implementation}\\
            \noalign{\smallskip}
            \includegraphics[height=2.5cm]{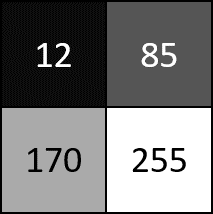} &
            \includegraphics[height=2.5cm]{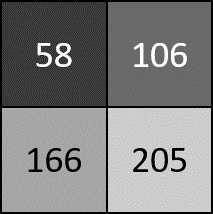}&
            \includegraphics[height=2.5cm]{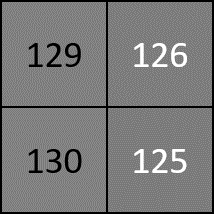} &
            \includegraphics[height=2.5cm]{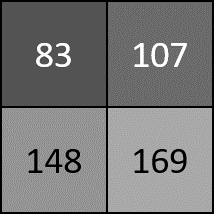} & \includegraphics[height=2.5cm]{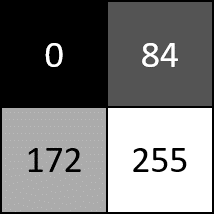}& 
            \includegraphics[height=2.5cm]{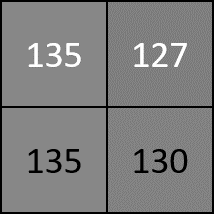}&
            \includegraphics[height=2.5cm]{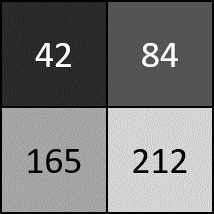}\\
            &\large best & \large worst & \large mean & \large best & \large worst & \large mean\\
            \large 'qasm\_simulator' & \multicolumn{3}{c|}{\large real backend} & \multicolumn{3}{c}{\large 'mitigation\_own'}\\
            \end{tabular}}
    \end{minipage}
    \caption{$2\times 2$ pixel gray value input image and resulting output images. We use the 'qasm\_simulator' with $8.192$ shots and in total $42$ executions on the five backends. We show the output images with the smallest (best) and highest relative difference (worst) as well as the mean of the $42$ executions and the five backends. The best results were obtained on backend 'ibmq\_manila'. The worst cases came from 'ibmq\_16\_melbourne'.}
	\label{fig:geng_alexander:image_out_all2x2}
\end{figure}

Relative differences in the range of $2-3\%$, in some executions also lower, can be achieved by using 'mitigation\_own'. 
Thus, with our adjustments and improvements, it is possible to reconstruct the input image. The outcomes with the smallest, highest, and mean relative difference are shown in Figure~\ref{fig:geng_alexander:image_out_all2x2}.

\subsection{Current maximal possible image sizes for simulator and NISQ quantum computer}
In \cite{co2021quantum}, an input image of size $32 \times 32$ was implemented. Following the same idea, we implement circuits for images of size $2^n\times2^n$, where $n\leq9$ (details as supplementary information in Section~\ref{section:supplementary}). Input pixel values are chosen randomly up to a size of $16\times16$. From $32\times32$ on, we use downscaled versions of the 8-bit gray value Lena image \cite{lena}, which originally has a size of $512\times512$ and is a standard test image in image processing.

We apply the steps from Sections~\ref{sec:practical_implementation}, \ref{sec:modification_mary}, and \ref{sec:extension} to check feasibility but do not numerically compare the output with the original input. Results are strongly influenced by noise as the image size is larger than $2\times 2$ and thus the associated circuit depth increases. This complicates the retrieval of an image enormously, since in a probabilistic model like FRQI the exact measurement of the probabilities for the individual states is crucial. The retrieved pixel values are all around $125$ using the five backends from above or the backend 'ibmq\_toronto' and do not reflect the input image at all. 

\begin{figure}[tb]
	\centering
	\begin{subfigure}[tb]{.24\linewidth}
	    \centering
        \includegraphics[width=0.9\textwidth]{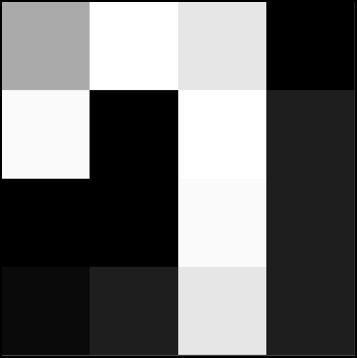}
        \caption{Input image}
        \label{fig:geng_alexander:image_in4x4}
    \end{subfigure}
    \begin{subfigure}[tb]{.24\linewidth}
        \centering
        \includegraphics[width=0.9\textwidth]{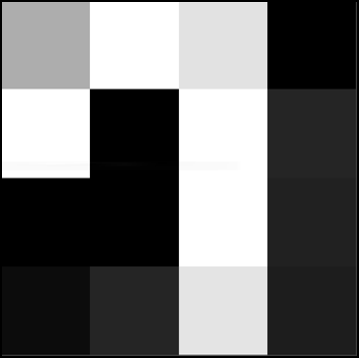}
        \caption{'qasm\_simulator'}
        \label{fig:geng_alexander:image_out_all_4x4_sim}
    \end{subfigure}
    \begin{subfigure}[tb]{.24\linewidth}
        \centering
        \includegraphics[width=0.9\textwidth]{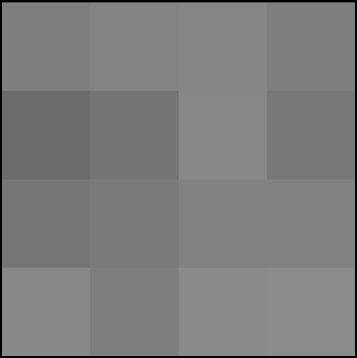}
        \caption{'ibmq\_toronto'}
        \label{fig:geng_alexander:image_out_all_4x4_real}
    \end{subfigure}
    \begin{subfigure}[tb]{.24\linewidth}
        \centering
        \includegraphics[width=0.9\textwidth]{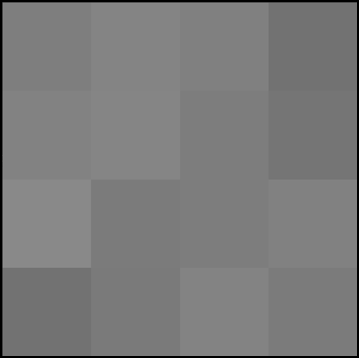}
        \caption{'mitigation\_own'}
        \label{fig:geng_alexander:image_out_all_4x4_mit}
    \end{subfigure}
	\caption{$4\times 4$ pixel downscaled gray value input image from the MNIST dataset \cite{deng2012mnist} and resulting output images. We use the MARY-implementation, $8.192$ shots for the 'qasm\_simulator', and 'ibmq\_toronto' as real backend. The results can only be improved slightly with the measurement error mitigation method.}
	\label{fig:geng_alexander:image_out_all_4x4}
\end{figure}

We further illustrate this in Figure~\ref{fig:geng_alexander:image_out_all_4x4}. Instead of the random image, we use a $4\times 4$ pixel downscaled gray value image from the MNIST dataset \cite{deng2012mnist}. This image shows a well recognizable digit seven. However, even with the MARY-implementation and when including error mitigation, the seven is no longer visible in the output image. 

Consequently, we can only recover images with a size up to $2\times2$ with FRQI on a real backend. Above that, the noise makes the recovery of an image impossible with the used implementation even when using measurement error mitigation.

\begin{figure}[tb]
	\centering
	\begin{subfigure}[tb]{.49\linewidth}
        \includegraphics[width=0.99\textwidth]{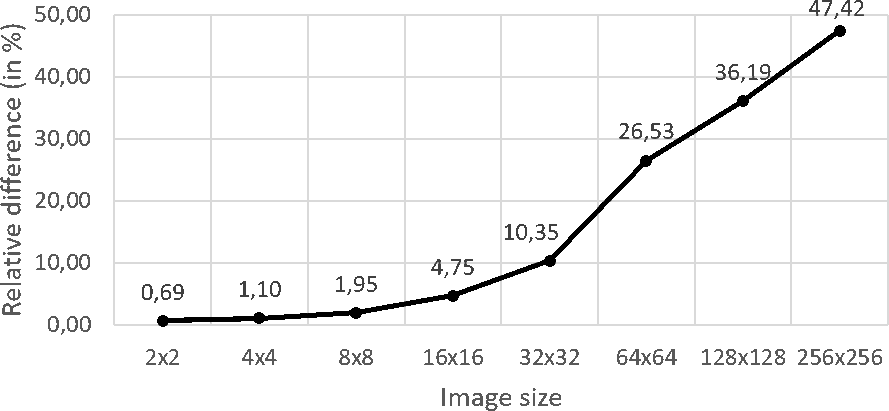}
        \caption{$8.192$ shots.}
        \label{fig:geng_alexander:extension_simulator_8192_shots}
    \end{subfigure}
    \begin{subfigure}[tb]{.49\linewidth}
        \includegraphics[width=0.99\textwidth]{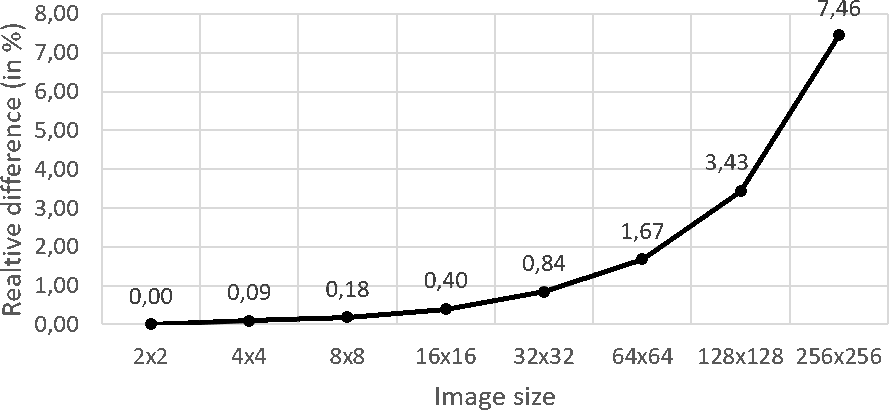}
        \caption{$10^6$ shots.}
        \label{fig:geng_alexander:extension_simulator_million_shots}
    \end{subfigure}
	\caption{Relative difference (in \%) between $image_{in}$ and $image_{out}$ for varying image sizes, 'qasm\_simulator' as backend and $8.192$ or $10^6$ shots.}
	\label{fig:geng_alexander:extension_simulator}
\end{figure}

\begin{figure}[tb]
	\centering
	\begin{subfigure}[tb]{.32\linewidth}
	    \centering
        \includegraphics[width=0.9\textwidth]{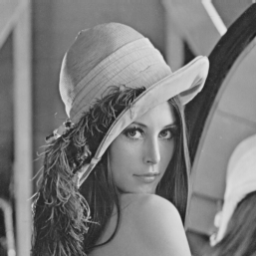}
        \caption{Input image}
        \label{fig:geng_alexander:image_out_sim_shots256_in}
    \end{subfigure}
    \begin{subfigure}[tb]{.32\linewidth}
        \centering
        \includegraphics[width=0.9\textwidth]{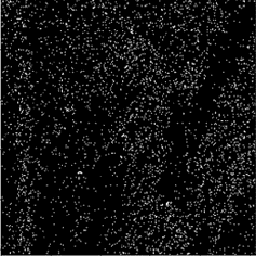}
        \caption{$8.192$ shots}
        \label{fig:geng_alexander:image_out_sim_shots256_8192}
    \end{subfigure}
    \begin{subfigure}[tb]{.32\linewidth}
        \centering
        \includegraphics[width=0.9\textwidth]{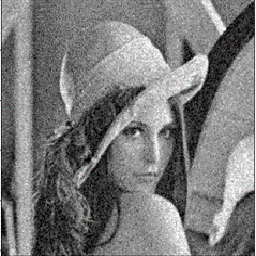}
        \caption{$10^6$ shots}
        \label{fig:geng_alexander:image_out_sim_shots256_1000000}
    \end{subfigure}
	\caption{$256\times 256$ pixel gray value downscaled Lena image \cite{lena} with 'qasm\_simulator' outcomes and varying number of shots.}
	\label{fig:geng_alexander:image_out_sim_shots256}
\end{figure}

Even with the noise free and fully connected 'qasm\_simulator', using $8.192$ shots for larger images results in images strongly deviating from the input image, as shown in Figure~\ref{fig:geng_alexander:extension_simulator_8192_shots}. 
Even a much higher number of shots
is no complete remedy (see Figure~\ref{fig:geng_alexander:extension_simulator_million_shots}). The visual differences for the downscaled $256\times 256$ Lena image are shown in Figure~\ref{fig:geng_alexander:image_out_sim_shots256}.

The observed deviation is related to the number of states that can occur for a given image size. For example in the MARY-implementation, we only increase the size of the MARY-gates without adding an ancilla qubit for images smaller than $32\times32$. Therefore, we have in total $2^{2n+1}$ possible states (like in the MCRY-implementation), where $2n+1$ is the number of required qubits. For images larger than or equal to $32\times32$, we have an ancilla qubit in addition to the positions qubits and the gray value qubit, so $2^{2n+2}$ possible states are conceivable. 

This results in more than a million possible states for images of size $512\times512$, since $20$ qubits are needed. Thus, accurate estimation of the distribution of possible states requires a sufficiently large sample. The number of possible states and the deviation from the input image grows exponentially as visualized in Figure~\ref{fig:geng_alexander:extension_simulator}.

\begin{figure}[tb]
	\centering
	\begin{subfigure}[tb]{\linewidth}
	    \includegraphics[width=0.99\textwidth]{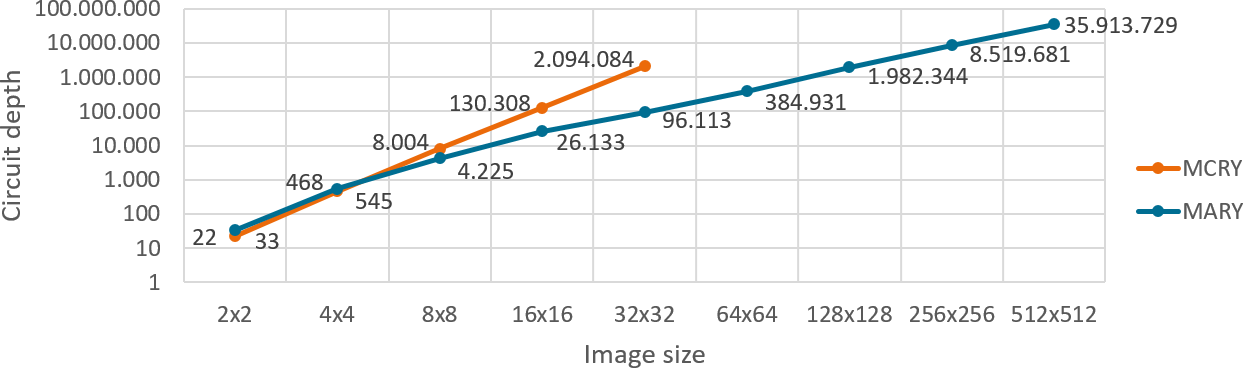}
	    \caption{'qasm\_simulator'}
	    \label{fig:geng_alexander:circuit_depth_sim}
	 \end{subfigure}
	 \begin{subfigure}[tb]{\linewidth}
	    \vspace{0.5cm}
	    \includegraphics[width=0.99\textwidth]{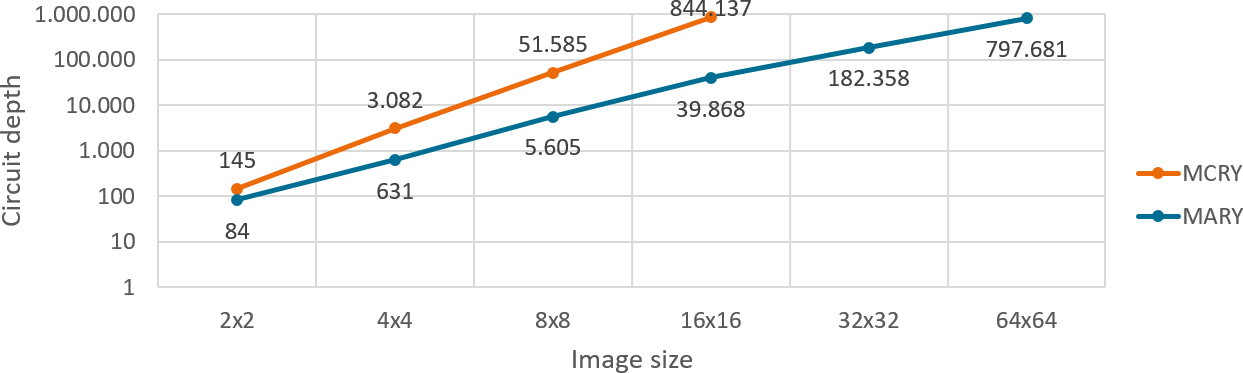}
	  \caption{'ibmq\_toronto'}
	  \label{fig:geng_alexander:circuit_depth_tor}
	 \end{subfigure}
	\caption{Circuit depth for varying image sizes and MCRY-/MARY-implementation. Values are mean values from $10$ observations and shown in logarithmic scale. Used backends: (a) 'qasm\_simulator', (b) 'ibmq\_toronto'.}
	\label{fig:geng_alexander:circuit_depth}
\end{figure}

The circuit depth increases exponentially, too. This is due to the entanglement of the position qubits with the gray value qubit and the associated exponential increase of the number of \CX-gates. See Figure~\ref{fig:geng_alexander:circuit_depth} for a visualization.

With 'qasm\_simulator', the circuit depth can be determined up to an image size of $32\times 32$ for MCRY- and $512\times 512$ for MARY-implementation without exceeding memory (see Figure~\ref{fig:geng_alexander:circuit_depth_sim}). However, actual execution is no longer possible from these image sizes onwards with $64$GB RAM available in the classical machine used for generating and storing the circuits.

For the backend 'ibmq\_toronto', the circuit depth allows calculation for image sizes up to $16\times 16$ for the MCRY- and $64\times 64$ for the MARY-implementation only, see Figure~\ref{fig:geng_alexander:circuit_depth_tor}. This is due to the fact that more gates are needed compared to the simulator. Additionally, backend specific things like coupling maps and errors are considered in the transpilation step which increases the memory needed. For the real backend, circuit depths vary due to changing errors, which is why we average over $10$ observations in Figure~\ref{fig:geng_alexander:circuit_depth_tor}.

The circuit depth is significantly higher for the MCRY-implementation which implies that more noise is accumulated in the process. Additionally, more memory has to be used which limits the executability of the MCRY-implementation.

All in all, we need more and more memory to generate the circuits for increasing image sizes. In the end, this is the limiting factor. The current possible image sizes that can be handled on 64GB RAM are shown in Table~\ref{tab:current_limits}. For larger images, the available memory is exceeded and the job aborts. If we focus on the outcomes, the finite sampling shot noise increases the relative difference for the 'qasm\_simulator'. The additional noise from gates and the environment further restricts the maximal image size for the real backend. In total, in spite of all suggested improvements, the reconstruction of the image with FRQI is only possible for $2\times 2$ images on the real backends.

\begin{table}
\caption{Current maximum executable and usable image sizes for MCRY- and MARY-implementation on 'qasm\_simulator' with $8.192$ shots and IBM's backend 'ibmq\_toronto' limited to $64$GB memory. The term executable refers to the possibility to run the algorithm without focusing on the outcomes. The term usable implies that the relative difference between input image and reconstructed image is less than $5\%$. }
\label{tab:current_limits}       
\begin{tabularx}{\textwidth}{lcccccc}
\hline\noalign{\smallskip}
 && \multicolumn{2}{c}{maximum executable image size} && \multicolumn{2}{c}{maximum usable image size}\\
\noalign{\smallskip}\hline\noalign{\smallskip}
Method && 'qasm\_simulator' & 'ibmq\_toronto' && 'qasm\_simulator' & 'ibmq\_toronto'  \\
\noalign{\smallskip}\hline\noalign{\smallskip}
MCRY  && $32\times32$  & $16\times16$ && $16\times16$ & $2\times2$ \\
MARY  && $256\times256$& $32\times32$ && $16\times16$ & $2\times2$\\
\noalign{\smallskip}\hline
\end{tabularx}
\end{table}

\begin{figure}[tb]
	\centering
	\begin{subfigure}[tb]{\linewidth}
	    \includegraphics[width=0.99\textwidth]{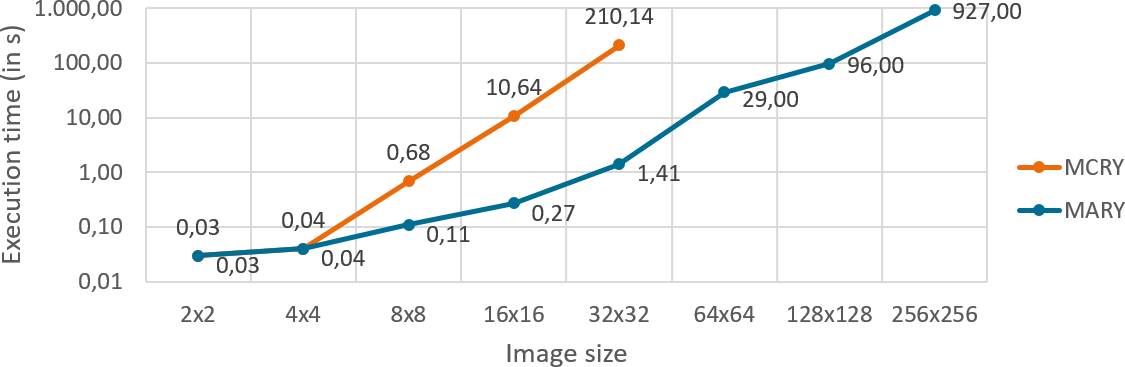}
	    \caption{'qasm\_simulator'}
	    \label{fig:geng_alexander:execution_time_sim}
	 \end{subfigure}
	 \begin{subfigure}[tb]{0.7\linewidth}
	    \vspace{0.5cm}
	    \includegraphics[width=0.99\textwidth]{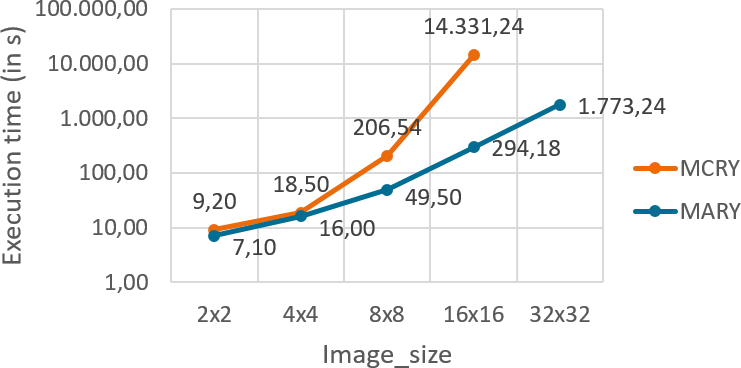}
	  \caption{'ibmq\_toronto'}
	  \label{fig:geng_alexander:execution_time_tor}
	 \end{subfigure}
	\caption{Execution time for varying image sizes and MCRY-/MARY-implementation. Values are mean values from $10$ observations and shown in logarithmic scale. Used backends: (a) 'qasm\_simulator' with $8.192$ shots, (b) 'ibmq\_toronto'.}
	\label{fig:geng_alexander:execution_time}
\end{figure}

Additionally, there are also significant differences in the execution times of the MCRY- and the MARY-implementation. This is shown in Figure~\ref{fig:geng_alexander:execution_time}. The higher circuit depth and number of gates in the MCRY-implementation result in higher execution times for both the 'qasm\_simulator' and IBM's backend 'ibmq\_toronto'. This difference is more pronounced for larger image sizes.

\section{Conclusion}\label{sect:conclusion}

In this paper, we investigate the practical use of one of the basic quantum image representations, namely FRQI. In particular, we determine the manageable image sizes both in terms of useable results and memory requirements. With exponentially increasing image size, the number of qubits increases only linearly. However, the number of possible states increases exponentially. As a result, even when using 'qasm\_simulator' and a fixed number of shots, it happens that pixel values of the output image cannot be distinguished from noise and therefore do not match those of the input image. 

Furthermore, the number of gates required for encoding an image with increasing size increases exponentially, too. Thus the influence of per-gate errors increases at the same speed. Our simplified implementation reduces this effect. It saves a large number of gates, reduces errors that way, and enables faster computation. The basic idea is to replace the error-prone multi-controlled gates by simplified versions needing less \CX-gates. With increasing image size, this strategy becomes more and more important. First, to obtain less confounded results. Second, to push the feasible image sizes to new frontiers.
\enlargethispage{\baselineskip}
All quantum image representations and all quantum algorithms involving multi-controlled operations can benefit from the presented simplified implementation.

\newpage

\bibliographystyle{unsrt}
\bibliography{main}  

\newpage
\section{Declarations}
\subsection{Funding}
This work was supported by the project AnQuC-2 of the Competence Center Quantum Computing Rhineland-Palatinate (Germany).
\subsection{Conflicts of interest/Competing interests}
The authors declare no competing interests.
\subsection{Availability of data and material}
All the data and simulations that support the findings are available from the corresponding author on request. 
\subsection{Code availability}
The jupyter notebooks used in this study are available from the corresponding author on request.
\newpage
\section{Supplementary information}\label{section:supplementary}
\begin{figure}[h]
	\centering
	\begin{subfigure}[tb]{\linewidth}
    	\includegraphics[width=\textwidth]{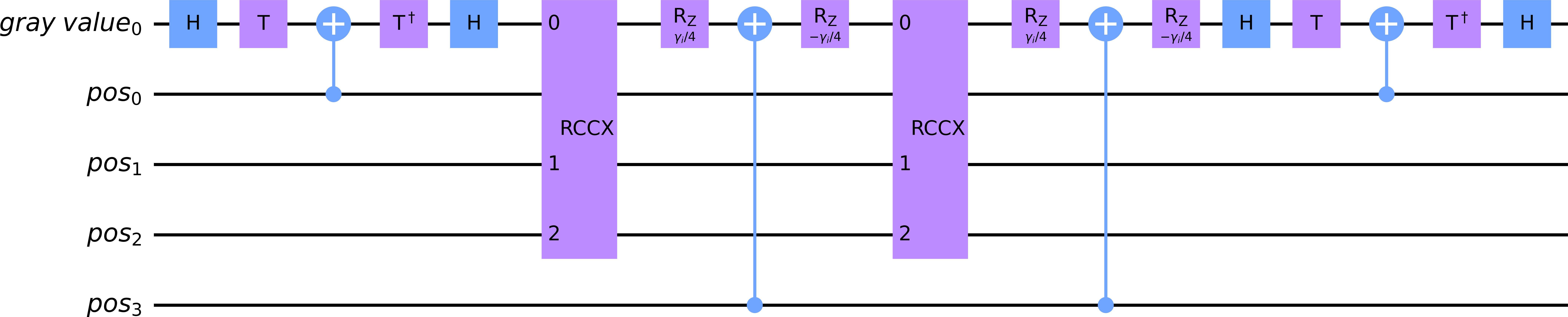}
    	\caption{Decomposed MARY5-gate for an arbitrary angle $\gamma_i=2\theta_i$ used to encode a $4\times4$ image by adding Hadamard and X-gates.}
    \end{subfigure}
    \begin{subfigure}[tb]{\linewidth}
    	\vspace{0.5cm}
    	\includegraphics[width=\textwidth]{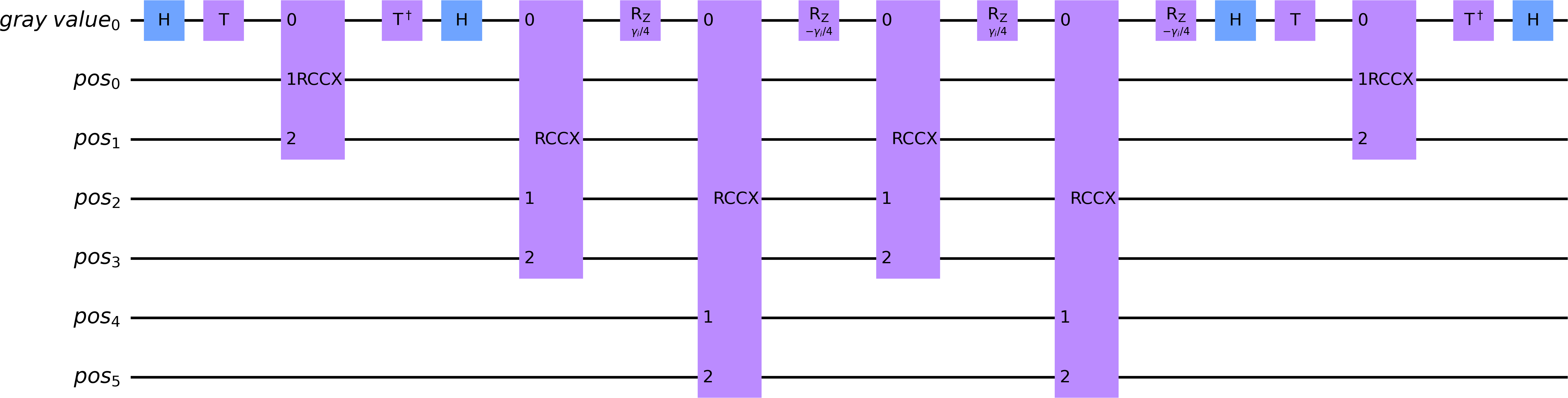}
    	\caption{Decomposed MARY7-gate for an arbitrary angle $\gamma_i=2\theta_i$ used to encode an $8\times8$ image by adding Hadamard and X-gates.}
    \end{subfigure}
    \begin{subfigure}[tb]{\linewidth}
    	\vspace{0.5cm}
    	\includegraphics[width=\textwidth]{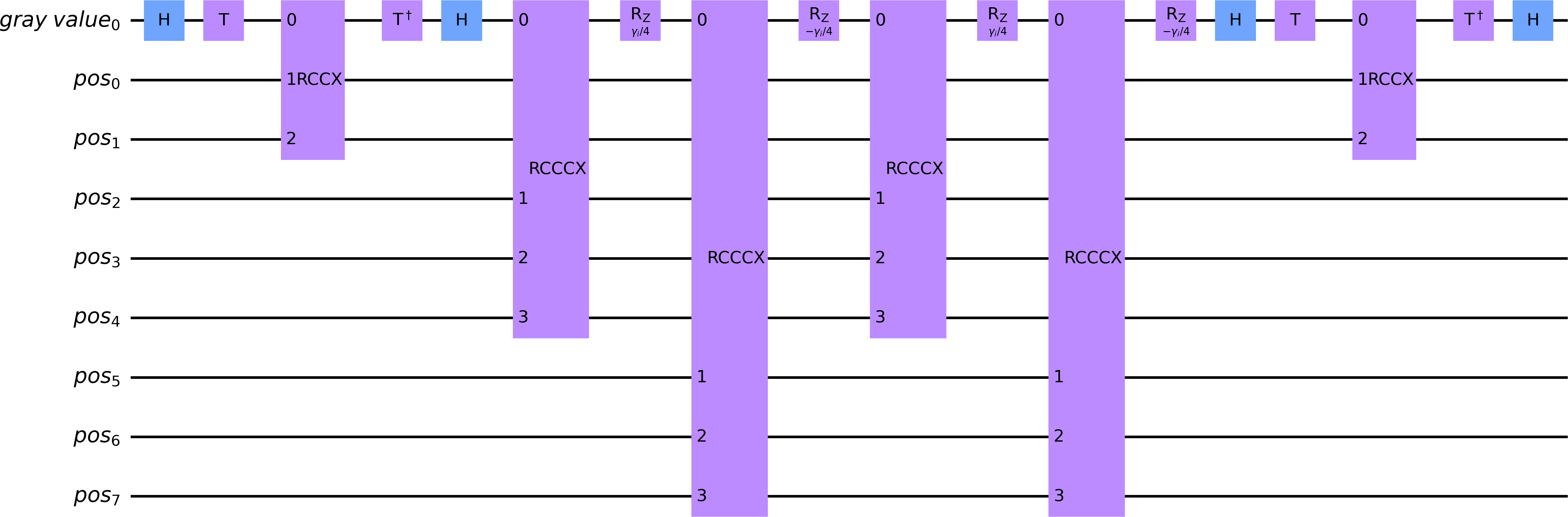}
    	\caption{Decomposed MARY9-gate for an arbitrary angle $\gamma_i=2\theta_i$ used to encode a $16\times16$ image by adding Hadamard and X-gates.}
    	\label{fig:geng_alexander:mary9_gate}
    \end{subfigure}
\end{figure}
\begin{figure}[tb]
    \ContinuedFloat
    \begin{subfigure}[tb]{\linewidth}
    	\includegraphics[width=\textwidth]{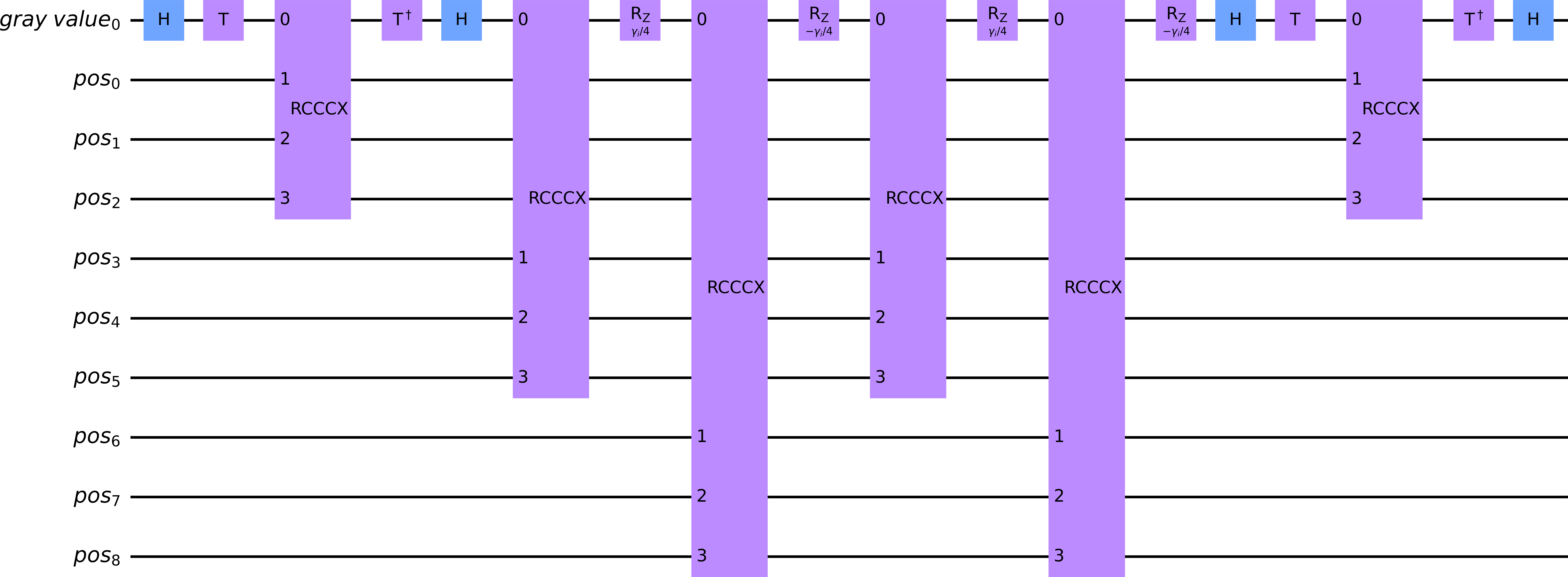}
    	\caption{Decomposed MARY10-gate for an arbitrary angle $\gamma_i=2\theta_i$ needed in our implementations for images with size $2^n\times 2^n$, where $n\in\{8,9,13\}$.}
    	\label{fig:geng_alexander:mary10_gate}
    \end{subfigure}
    \begin{subfigure}[tb]{\linewidth}
        \vspace{0.5cm}
    	\includegraphics[width=\textwidth]{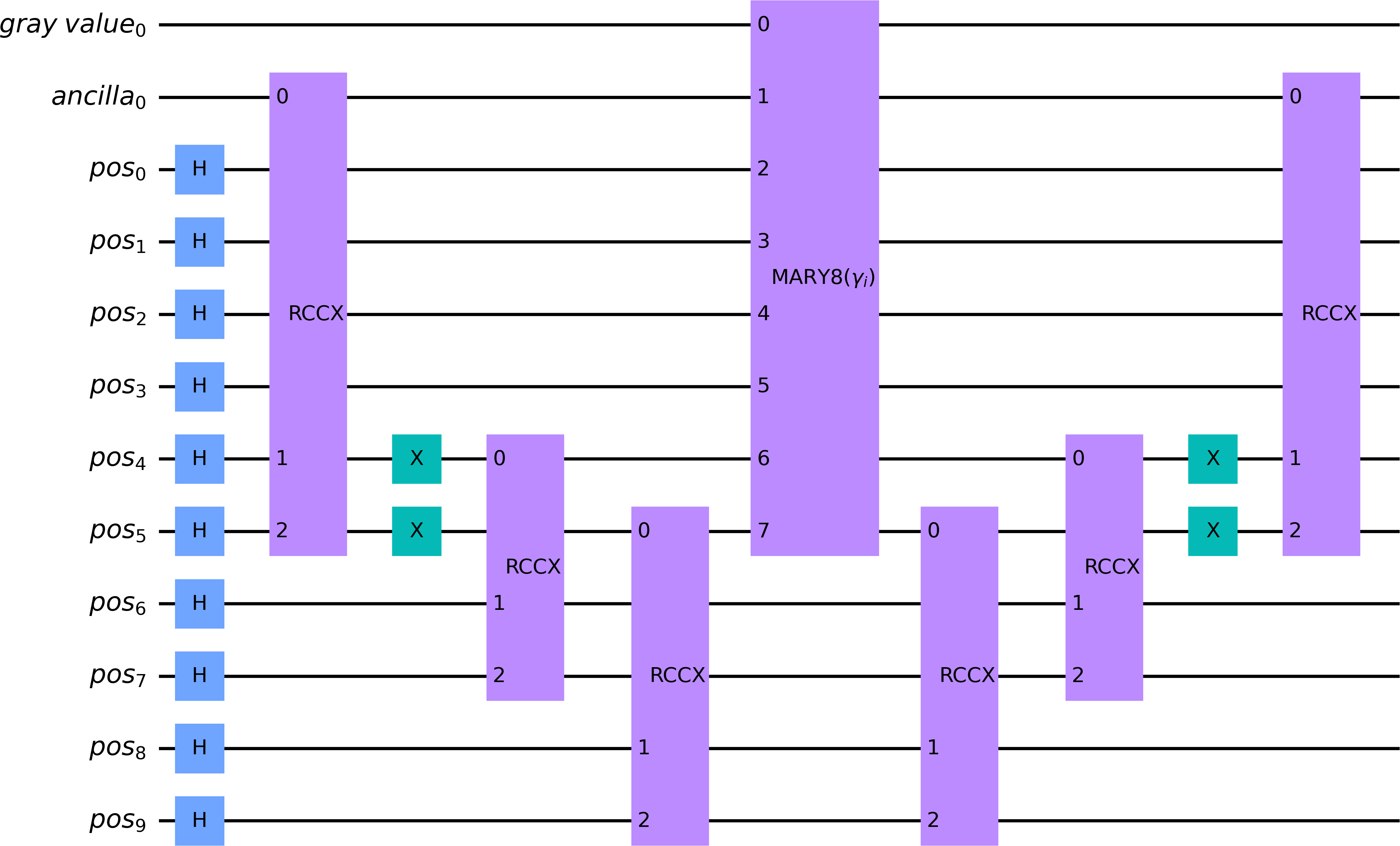}
    	\caption{Circuit for one angle $\gamma_i=2\theta_i$ of a $32\times32$ image. X-gates after the Hadamard gates are not shown here. The decomposed MARY8-gate is shown in Figure~\ref{fig:geng_alexander:mary8_gate}.}
    \end{subfigure}
\end{figure}
\begin{figure}[tb]
    \ContinuedFloat
    \begin{subfigure}[tb]{\linewidth}
    	\vspace{0.5cm}
    	\includegraphics[width=\textwidth]{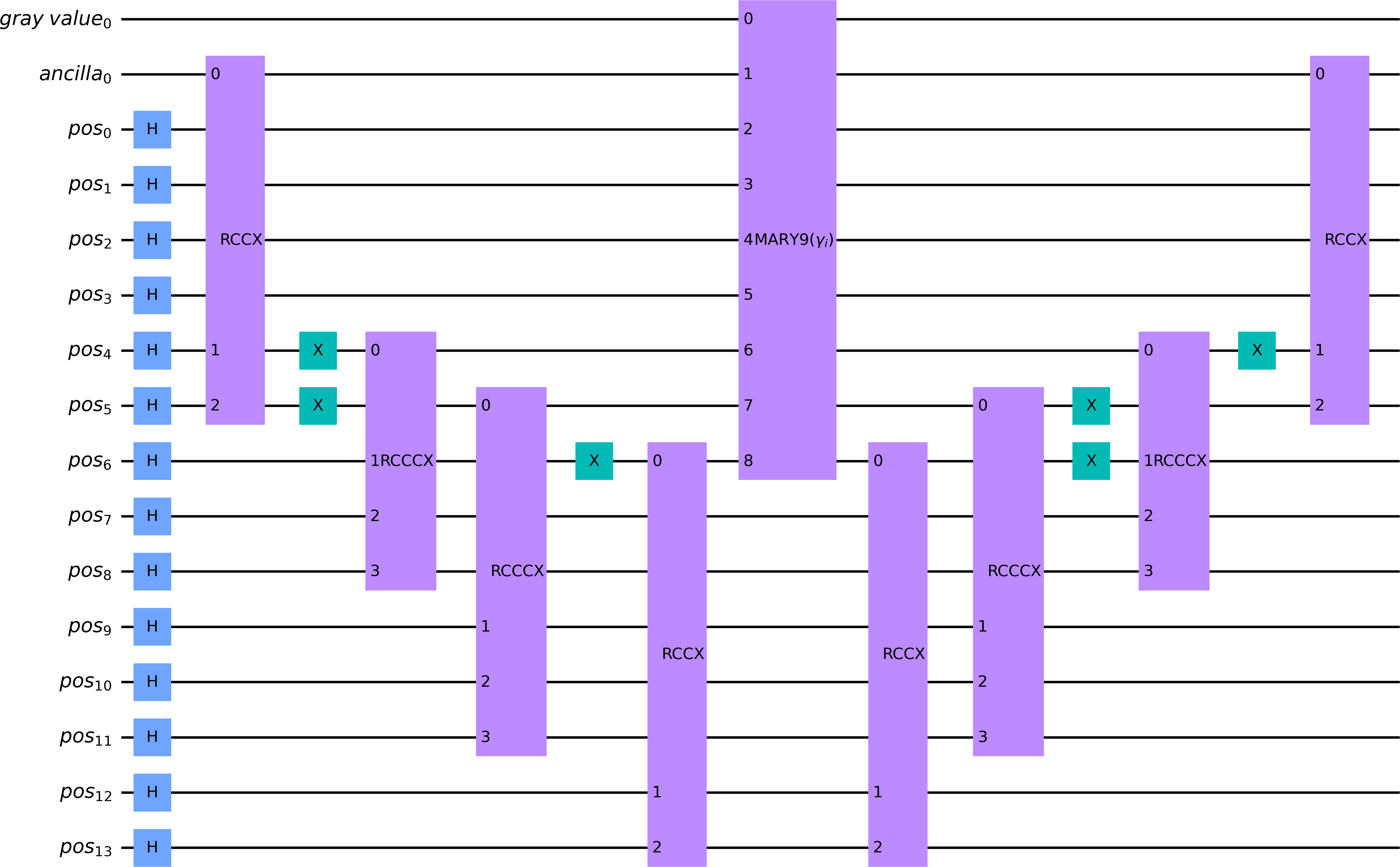}
    	\caption{Circuit for one angle $\gamma_i=2\theta_i$ of a $128\times128$ image. X-gates after the Hadamard gates are not shown here. The decomposed MARY9-gate is shown in Figure~\ref{fig:geng_alexander:mary9_gate}.}
    \end{subfigure}
    \begin{subfigure}[tb]{\linewidth}
    	\includegraphics[width=\textwidth]{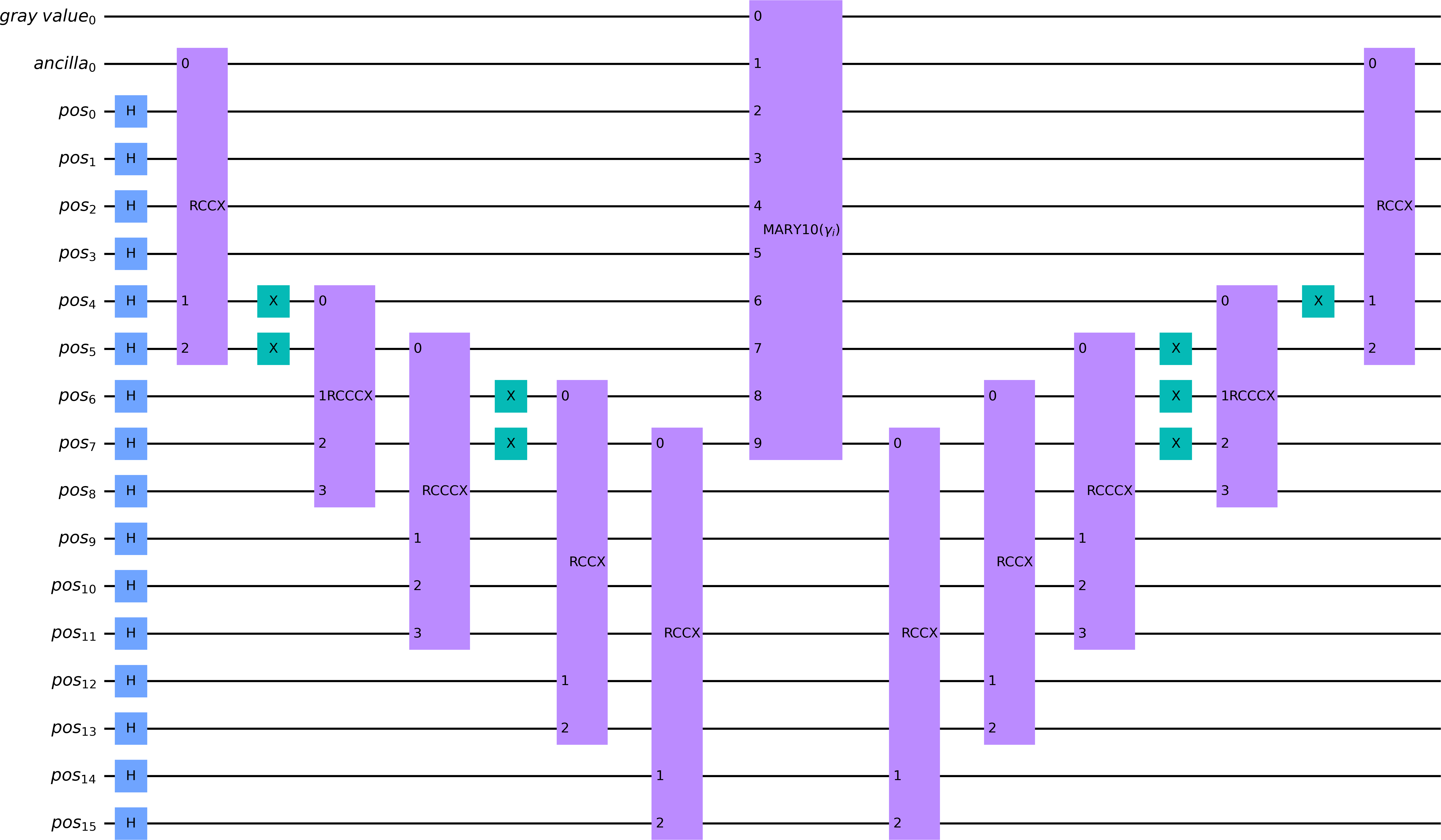}
    	\caption{Circuit for one angle $\gamma_i=2\theta_i$ of a $256\times256$ image. X-gates after the Hadamard gates are not shown here. The decomposed MARY10-gate is shown in Figure~\ref{fig:geng_alexander:mary10_gate}.}
    \end{subfigure}
\end{figure}
\begin{figure}[tb]
    \ContinuedFloat
    \begin{subfigure}[tb]{\linewidth}
        \vspace{0.5cm}
    	\includegraphics[width=\textwidth]{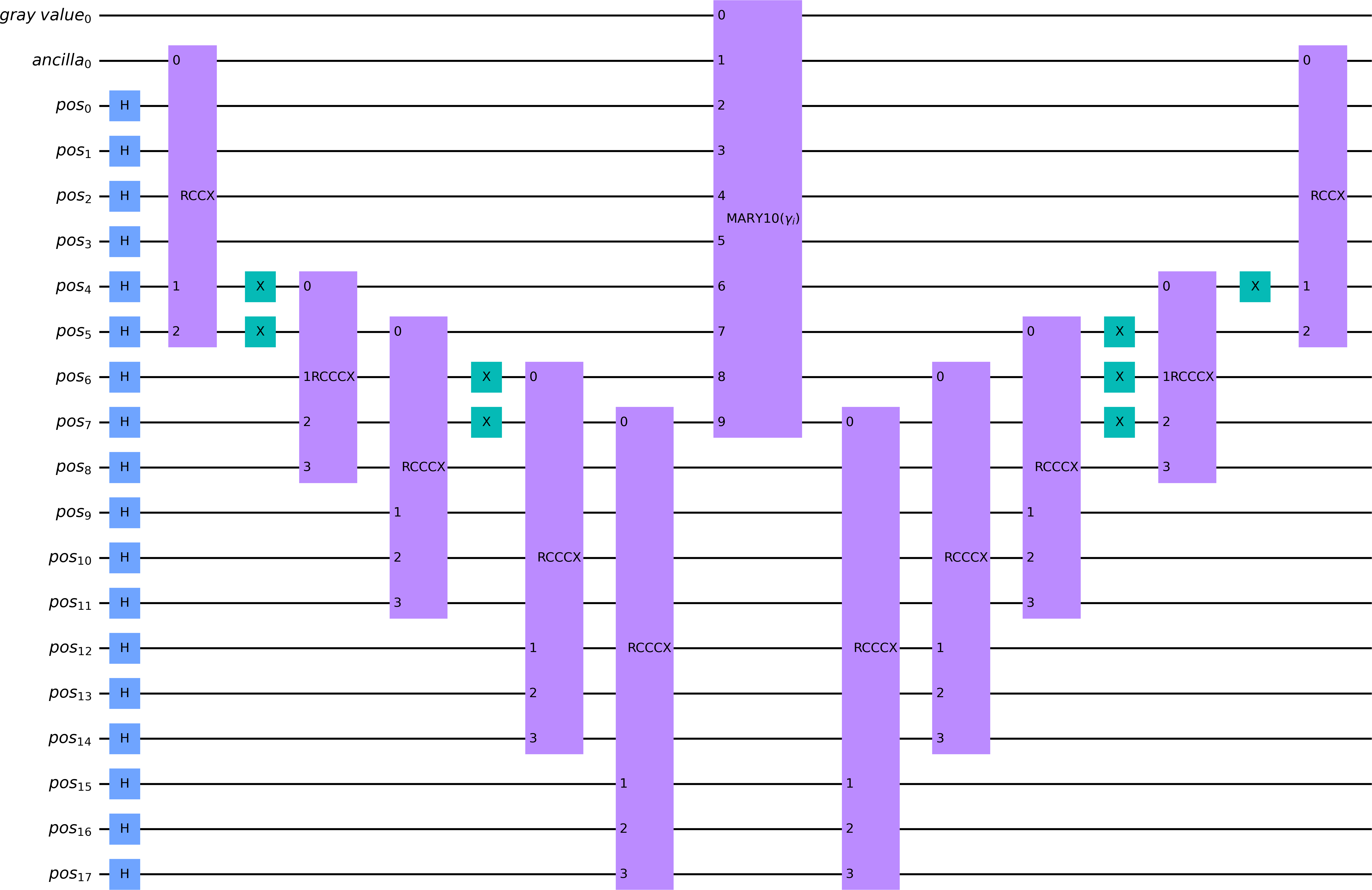}
    	\caption{Circuit for one angle $\gamma_i=2\theta_i$ of a $512\times512$ image. X-gates after the Hadamard gates are not shown here. The decomposed MARY10-gate is shown in Figure~\ref{fig:geng_alexander:mary10_gate}.}
    \end{subfigure}
\end{figure}
\begin{figure}[tb]
    \ContinuedFloat
    \begin{subfigure}[tb]{\linewidth}
    	\includegraphics[width=\textwidth]{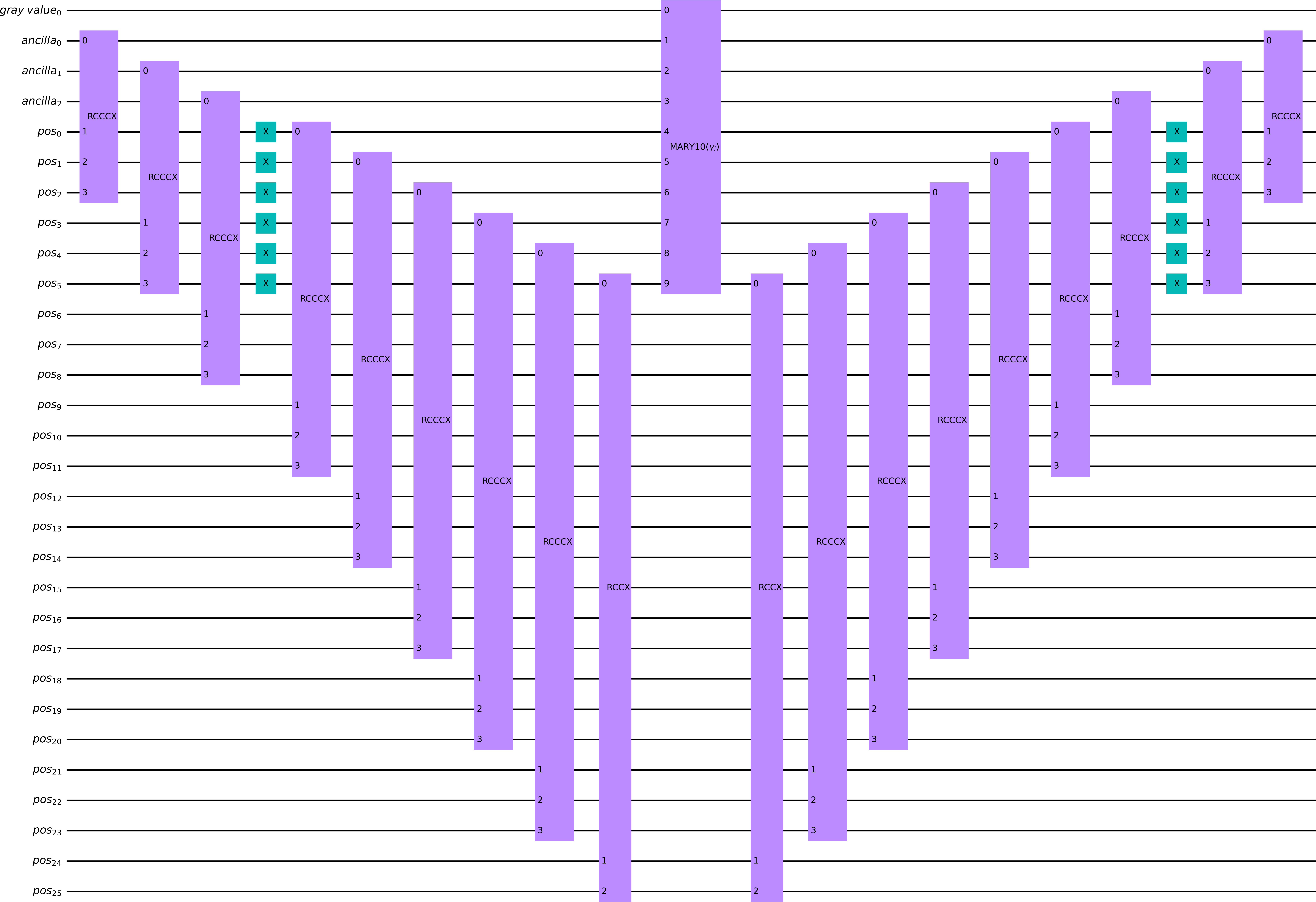}
    	\caption{Circuit for one angle $\gamma_i=2\theta_i$ of a $8.192\times8.192$ image. X-gates and Hadamard gates are not shown here. The decomposed MARY10-gate is shown in Figure~\ref{fig:geng_alexander:mary10_gate}.}
    	\label{fig:geng_alexander:circuit8192}
    \end{subfigure}
    \caption{Implemented circuits for varying image sizes shown via Qiskit's 'draw' function \cite{qiskit_short} with matplotlib output variant 'mpl'.}
\end{figure}

\end{document}